\documentclass[10pt]{iopart}
\usepackage{graphicx,epsfig}
\usepackage{bm}
\usepackage{iopams}
\newcommand{\ba}{\begin{eqnarray}}
\newcommand{\ea}{\end{eqnarray}}
\newcommand{\bse}{\numparts}
\newcommand{\ese}{\endnumparts}
\newcommand{\ACal}{{\cal{A}}}
\newcommand{\B}{{\cal{B}}}

\newcommand{\C}{{\cal {C}}}
\newcommand{\DD}{{\cal {D}}}

\newcommand{\W}{{\cal {W}}}
\newcommand{\bbq}{\begin{quote}}
\newcommand{\eeq}{\end{quote}}
\newcommand{\RR}{{}^3{\cal{R}}}

\newcommand{\T}{{}^3{\cal{T}}}

\newcommand{\EE}{{\cal{E}}}
\newcommand{\FF}{{\cal{F}}}

\newcommand{\VV}{{\cal{V}}}

\newcommand{\MM}{{\cal{M}}}
\newcommand{\QQ}{{\cal{Q}}}

\newcommand{\rhoav}{\langle\rho\rangle}

\newcommand{\Thetaav}{\langle\Theta\rangle}

\newcommand{\RRav}{\langle\RR\rangle}

\newcommand{\FFav}{\langle\FF\rangle}

\newcommand{\Aav}{\langle A\rangle}
\newcommand{\Bav}{\langle B\rangle}
\newcommand{\Fav}{\langle F\rangle}

\newcommand{\Dth}{\delta^{(\Theta)}}

\newcommand{\dd}{{\rm{d}}}
\newcommand{\rtv}{r_{\rm{tv}}}

\begin{document}


\title[On spatial averaging in Lema\^{\i}tre--Tolman--Bondi dust models. Part I.]{On spatial volume averaging in Lema\^{\i}tre--Tolman--Bondi dust models. Part I:  back reaction, spacial curvature and binding energy.} 
\author{ Roberto A. Sussman$^\ddagger$}
\address{
$^\ddagger$Instituto de Ciencias Nucleares, Universidad Nacional Aut\'onoma de M\'exico (ICN-UNAM),
A. P. 70--543, 04510 M\'exico D. F., M\'exico. }
\ead{sussman@nucleares.unam.mx}
\date{\today}
\begin{abstract} We provide a comprehensive analytic study (rigorous and qualitative) of the conditions for the existence of a a positive kinematic back reaction term $\QQ>0$, in the context of Buchert's scalar averaging formalism applied to spherically symmetric Lema\^{\i}tre--Tolman--Bondi (LTB) dust solutions in which averaging domains are given as spherical comoving regions containing a symmetry center. We introduce proper volume and quasi--local average functionals and functions in order to examine the conditions for $\QQ\geq 0$, and in the process we also explore the relation between back reaction, spatial curvature and binding energy for a wide variety of LTB configurations. The back reaction term is positive for all ``hyperbolic'' regular domains with negative spatial curvature, either in the full radial range or in the radial asymptotic range.  This result is also valid if these domains contain an inner ``elliptic'' region with positive curvature undergoing local collapse.  For some cases in which positive spatial curvature decreases asymptotically, the conditions for a positive back reaction can still be met but seem to be more restrictive. Since $\QQ>0$ is a necessary condition for a positive ``effective'' acceleration that would mimic the effect of dark energy (in the context of Buchert's formalism), we examine this issue in LTB models in a follow up paper (part II).                 
\end{abstract}

\maketitle
\section{Introduction.}

The possibility that observations could be influenced by diferent ways of coarse graining and averaging of dust inhomogeneities provides a number of popular alternative explanations for cosmic acceleration found in the literature~\cite{InhObs1,InhObs2,InhObs3,InhObs4,InhObs5,InhObs6}. Among these theoretical proposals, the spatial averaging formalism developed by  Buchert~\cite{buchert} and co--workers considers the ``effective'' acceleration that mimics a cosmological constant, which might arise from the so--called ``back--reaction'' terms that emerge by rewriting scalar evolution equations in terms of spatial averages of matter--energy density and the expansion scalar. See \cite{ave_review} for a comprehensive review of this formalism and \cite{IW} for further discussion. For alternative proposals on averaging inhomogeneities, see \cite{zala,colpel}. 

Unfortunately, whether based on Buchert's formalism or not, it is technically challenging to calculate in practice coarse grained observational parameters or back--reaction terms for general non--linear inhomogeneus or ``realistic'' conditions. Hence, besides perturbative approach~\cite{pert} and idealized spacetimes~\cite{wiltshire}, proposals that examine cosmic acceleration without dark energy have been discussed or tested for the spherically symmetric Lema\^\i tre--Tolman--Bondi (LTB) class of dust solutions~\cite{InhObs1,InhObs2,InhObs5,ltbmods,LTBave1,LTBave2,LTBave3}. These models~\cite{LTB,kras} are simple, but general enough, inhomogeneous spacetimes and so are ideal to test the effects of inhomogeneity.  See \cite{ltbstuff1,ltbstuff2, ltbstuff3, ltbstuff,suss02} for further discussion on regularity of LTB models and \cite{kras} for a comprehensive review.   

Several articles among the references in \cite{InhObs1,InhObs2,InhObs5,ltbmods,LTBave1} have already considered the application of Buchert's formalism to LTB models.  More recently, Paranjape and Singh~\cite{LTBave2} utilized asymptotic or late time approximations, while Chuang, Gu and Hwang \cite{LTBave3} relied on particular cases of exact LTB solutions. These articles apparently signal that a positive effective acceleration can occur under certain conditions (negative curvature and low density). In the present article and its continuation (part II) we enhance and complement this existing literature by looking at this theoretical issue in in full general analytic form, without specializing for particular LTB models, and without perturbations or approximations. Specifically: we examine in detail sufficient conditions for a non--negative back--reaction term, $\QQ\geq 0$, which become necessary conditions for the existence of a positive effective  acceleration $A_{{\rm{eff}}}\geq 0$ (applied in part II to those cases previously found in this paper to comply with $\QQ\geq 0$). We have summarized these results in Table 1, for the benefit of those readers who wish to see them before going into the technical detail. The relation between back--reaction, spatial curvature and a binding energy functional, which can be defined in a covariant manner~\cite{sussQL}, appears in the process of discussing the conditions for $\QQ\geq 0$. This relation, summarized  in section 14,  provides a solid, though model--dependent, support for the arguments suggested by Wiltshire~\cite{wiltshire2} on back--reaction, quasi--local energy and binding energy.   
   
We summarize now the section by section contents of the article. We present in section 2 the basic parameters and properties of LTB models. In section 3 we examine the natural and covariant time slicing afforded by the normal geodesic 4--velocity, discussing geometric properties of the hypersurfaces $\T(t)$ orthogonal to $u^a$, which provide the spatial slices in which the proper volume averages associated with Buchert's formalism will be defined and computed.  Section 4 provides a rigorous definition of the proper volume spatial averaging functional, as well as its associated local valued function, while in section 5 we introduce the quasi--local average functional and function, which will be needed in subsequent sections. The evolution equations for averaged scalars in Buchert's formalism, including the back--reaction term $\QQ$ and its associated effective acceleration $A_{{\rm{eff}}}$, are given in section 6, while in section 7 we present the generic sufficient conditions for $\QQ\geq 0$. We discuss the  conditions for the existence of zeros of radial gradients of scalars  (``turning values'' or TV's) in section 8. The sufficient conditions for $\QQ\geq 0$ are then applied to domains in various LTB configurations: ``parabolic'' (section 9), ``hyperbolic'' (section 10), ``elliptic'' with open and closed topology (sections 11 and 12) and mixed configurations in which an inner elliptic region (enclosing a center) is surrounded by a hyperbolic or parabolic exterior (section 13). In section 14 we provide an overview and final discussion on the interrelation between back reaction, spatial curvature and binding energy. The discussion on properties of LTB models given in sections 2 and 3 is complemented with a summary of regularity conditions (Appendix A), the relation between radial coordinate and proper radial length  (Appendix B) and the fluid flow evolution equations~\cite{ellisbruni89,BDE,1plus3,LRS} (Appendix C). Appendices D and E provide the detailed proof of Propositions 7 and 8. 

The most relevant result in sections 9--13 is the fact that averaging domains always exist in which $\QQ\geq 0$ holds for all regular LTB models whose radial asymptotic range has, either negative spatial curvature ($\RR<0$ hyperbolic), or radially decaying positive curvature ($\RR>0$ elliptic models with open $\mathbb{R}^3$ topology) with sufficiently large gradients $\RR'<0$. In a follow up paper (part II) we explore conditions for which $A_{{\rm{eff}}}\geq 0$ can hold for those configurations in which $\QQ\geq 0$ holds, leading to a robust analytic proof on the compatibility between specific LTB models and the existence of a positive effective acceleration in the context of Buchert's formalism.     
   
\section{LTB dust spacetimes.}

Lemaitre--Tolman--Bondi (LTB) dust models~\cite{LTB,kras} are the spherically symmetric solutions of Einstein's equations characterized by the LTB line element and the energy--momentum tensor
\ba ds^2=-c^2dt^2+\frac{R'{}^2}{\FF^2}\,dr^2+R^2(d\theta^2+ 
\sin^2\theta d\phi^2),\label{LTB1}\\
T^{ab} = \rho\,c^2\,u^a\,u^b,\qquad u^a=\delta^a_0,\label{Tab}\ea
where $R=R(t,r)$,\,\ $R'=\partial R/\partial r $,  $\FF=\FF(r)\geq 0$ and $\rho=\rho(t,r)$ is the rest--mass energy--density. The field equations reduce to
\ba \dot R^2 &=& \frac{2M}{R} +(\FF^2-1),\label{fieldeq1}\\
 2M' &=& \kappa \rho \,R^2 \,R',\label{fieldeq2}\ea
where $\kappa=8\pi G/c^2$,\, $M=M(r)$ and $\dot R=u^a\nabla_a R=\partial R/\partial (ct)$. 

The basic kinematic parameters in LTB models are the expansion scalar, $\Theta=\nabla_au^a$, and the spatial trace--less shear tensor $\sigma_{ab}=h_a^ch_b^d\nabla_{(c}u_{d)}-(\Theta/3)h_{ab}$ 
\begin{equation}\Theta = \frac{2\dot R}{R}+\frac{\dot R'}{R'},
\label{Theta1}\end{equation}
\begin{equation}  
\sigma^{ab} = \Sigma\,\Xi^{ab}\quad \Rightarrow\quad \Sigma =\frac{1}{6}\,\sigma_{ab}\,\Xi^{ab}= -\frac{1}{3}\left(\frac{\dot R'}{R'}-\frac{\dot R}{R}\right),
\label{Sigma1}\end{equation}
where $h_{ab}=u_au_b+g_{ab}$,\, $\Xi^{ab}\equiv h^{ab}-3\chi^{a}\chi^b$ and $\chi^a=\sqrt{h^{rr}}\,\delta^a_r$ is the unit vector orthogonal to $u^a$ and to the 2--spheres orbits of SO(3). Another important covariant object is the electic Weyl tensor $E^{ab}=u_cu_dC^{abcd}$
\ba  
 E^{ab} = \EE\,\Xi^{ab}\quad \Rightarrow\quad \EE =-\frac{1}{6}\,\EE_{ab}\,\Xi^{ab}= -\frac{\kappa}{6}\rho+\frac{M}{R^3},
\label{EE1} \ea 
Notice that both $\sigma_{ab}$ and $E_{ab}$ can be described by the single scalar functions $\Sigma,\,\EE$ in a covariant manner. We provide in Appendix C the ``fluid flow'' evolution equations for the covariant scalars $\rho,\,\Theta,\,\Sigma$ and $\EE$.  

The term $\FF^2-1$ in (\ref{fieldeq1}) is often interpreted as a ``binding energy'' for comoving dust layers~\cite{kras,ltbstuff3,ltbstuff}, as (\ref{fieldeq1}) is analogous to a Newtonian energy equation. It is important to remark that all quantities in (\ref{fieldeq1})-(\ref{fieldeq2}) are are invariant scalars for LTB models: $M$ is  the conserved ``quasi--local'' mass of Misner and Sharp~\cite{MSQLM,hayward1,hayward2,szab}, $R$ is the ``area distance'', while $\FF$ can be related in a covariant manner (in spherical symmetry) to a binding energy functional (see section 14 and \cite{sussQL}). 

It is common usage in the literature to classify LTB models according to a ``kinematic equivalence class'', based on the sign of $\FF^2-1$, which determines the existence of a zero of $\dot R^2$ in (\ref{fieldeq1}).  If the given sign holds in the whole regularity domain, we have
\bse\ba \FF^2-1=0,\quad \hbox{or:}\quad \FF=1,\qquad\qquad \hbox{parabolic models},\label{parab}\\
\FF^2-1\geq 0,\quad \hbox{or:}\quad \FF\geq 1, \qquad\qquad \hbox{hyperbolic models},\label{hyperb}\\
\FF^2-1\leq 0,\quad \hbox{or:}\quad -1\leq\FF\leq 1, \qquad \hbox{elliptic models}\label{elliptic},
\ea\ese
with the equal sign above holding only at a symmetry center. In general, it is possible to consider $\FF^2-1$ changing sign in a given radial range, defining LTB models that contain hyperbolic or elliptic domains or ``regions'' (see \cite{kras,ltbstuff3,ltbstuff}). 

The standard technique to deal with LTB dust models is to solve the Friedman--like field equation (\ref{fieldeq1}) for suitably prescribed functions $M(r)$ and $\FF(r)$ (the latter commonly as a function $E=\FF^2-1$), using then these solutions to find the remaining relevant quantities that may be required for a specific problem. This procedure has lead to analytic solutions (mostly implicit or parametric) that are well known and have been used abundantly in the literature (see \cite{kras} for a comprehensive review). A different approach to study the dynamics of these models is the ``fluid flow'' or covariant ``1+3'' decomposition of Ehlers, Ellis, Bruni, Dunsbury and van Ellst~\cite{ellisbruni89,BDE,1plus3,LRS} (see Appendix C).  Standard regularity conditions for these models are discussed in \cite{ltbstuff1,ltbstuff2,ltbstuff3,ltbstuff,suss02} and summarized in Appendices A and B.

\section{Covariant time slicing and geometry of space slices.}

The normal comoving 4--velocity in (\ref{LTB1}) provides a covariant time slicing in which the space slices are the hypersurfaces $\T(t)$, orthogonal to $u^a$, and marked by an arbitrary but fixed $t=t_0$.  Since we will be considering integral functions and functionals along the $\T(t_0)$, it is important to provide a basic discussion on the geometric properties of these slices (see Appendices A and B for further discussion). 

The metric, proper volume element and 3--dimensional Ricci scalar associated with the $\T(t_0)$ are
\ba h_{ab}=g_{ab}+u_a u_b=g_{ij}\delta^i_a\delta^j_b,\quad i,j=r,\theta,\phi,\label{spmetric}\\
\dd \VV_p=\sqrt{{\rm{det}}(h_{ab})}\,\dd^3x=\FF^{-1}\,R^2 R'\,\sin^2\theta\,\dd r\,\dd\theta\,\dd\phi,\label{dV}\\
\RR = -\frac{2\,[(\FF^2-1)\,R]\,'}{R^2R'} =-\frac{-2\FF\FF'}{R'R}-\frac{2\,(\FF^2-1)}{R^2},\label{3Ricci1}\ea
where $R=R(t_0,r),\,R'=R'(t_0,r)$.  We will assume henceforth the existence of (at least) one regular symmetry center (see Appendix A) marked by $r=0$, so that $R(t,0)=\dot R(t, 0)=0$ for all $t$ and $r\geq 0$ (hypersurfaces $\T$ homeomorphic to $\mathbb{S}^3$ have a second symmetry center at $r=r_c$). 

Each $\T(t_0)$ is a warped product $\T=\chi(t_0, r)\times_R \mathbb{S}^2(\theta,\phi)$, where the warping function is $R(t_0,r)\geq 0$, the fibers are concentric 2--spheres $\mathbb{S}^2$ with surface area $4\pi R^2(t_0,r)$, while the leaves $\chi(t_0,r)$ are ``radial rays'' or curves of the form $[ct_0, r,\theta_0,\phi_0]$, with $\theta_0,\phi_0$ constants. Since the rays at each $\T(t_0)$ are isometric to each other, every scalar $A$ function in every $\T(t_0)$ is equivalent to a real valued function $A:\chi(t_0,r)\to\mathbb{R}$ that corresponds to one of the functions in the one--parameter family $A(ct_0,r)$. Evidently, for time dependent scalars we will have a different radial dependence at different $\T(t_0)$, while scalars like $M(r)$ or $\FF(r)$ have identical radial dependence in all $\T(t_0)$. 

It is important to mention that the radial coordinate in (\ref{LTB1}) and (\ref{spmetric}) has no inherent covariant meaning. However, as we show in Appendix B, as long as standard regularity conditions hold, the proper length along radial rays (which are spatial geodesics of these metrics) is a monotonically increasing function of $r$, and so the dependence of scalars on $r$ at ant $\T(t)$ is qualitatively analogous to their dependence on the proper length.  Since it is evident that all time dependent quantities defined on the $\T(t)$ depend on $t$ as a fixed parameter, we will henceforth omit expressing this dependence explicitly. Unless it is needed for clarity, we will use the symbol $A(r)$ instead of $A(t,r)$.

The hypersurfaces $\T(t)$ can be classified in terms of the sign of the spatial scalar curvature $\RR$ in (\ref{3Ricci1}), and the ``kinematic class'' given by the sign of $\FF^2-1$ as in (\ref{parab})--(\ref{elliptic}), which identifies the $\T(t)$ (or regions of them) as slices of parabolic, hyperbolic or elliptic models. Given the existence of (at least) one symmetry center, the topology (homeomorphic class) of the $\T(t)$ are

\begin{itemize}

\item {\underline{$\T(t)$ homeomorphic to $\mathbb{R}^3$}}. There is only one symmetry center, at $r=0$, hence we must have $R'>0$ everywhere. Notice that this also follows from demanding absence of shell crossings (see Appendix A). From the proper length definition (\ref{xidef}) and condition (\ref{RrF}) in Appendix B, we must have $\FF> 0$ for these cases, which means that this topology is compatible with regions or models of all kinematic classes (hyperbolic, parabolic or elliptic). As a consequence of (\ref{Rrxi}) and (\ref{sz2}), if $\xi\to\infty$ then $R\to \infty$, though it is possible to have $R\to$ constant if $\FF\to 0$ asymptotically. Notice that (\ref{Rrxi})  implies that all $\T(t)$ have the same topology.  

\item {\underline{$\T(t)$ homeomorphic to $\mathbb{S}^3$}}. There are two symmetry centers, at $r=0$ and $r=r_c$. Since $R(t,0)=R(t,r_c)=0$ for all $t$, then there must be a ``turning value'' $r=\rtv$ of $R$, so that $R'(\rtv)=0$ where $0<\rtv<r_c$.  Since the regularity conditions (\ref{RrF}) and (\ref{Rrxi}) require  $\FF(\rtv)=0$, all regular models whose $\T(t)$ have this topology must be elliptic (though elliptic models can also have $\T(t)$ homeomorphic to $\mathbb{R}^3$). See section 12 and Appendices A, B and E.           

\end{itemize}

The interrelation between kinematic class (sign of $\FF^2-1$ in (\ref{parab})--(\ref{elliptic})), radial profiles of $\FF$, scalar curvature $\RR$ in (\ref{3Ricci1}) and topology will be discussed for each case in sections 8--13.

\section{Proper volume average functionals and functions.}

Each regular
\footnote{We are assuming that the $\T(t)$ are fully regular, which is not true when a curvature singularity arises at a given $t$. We discuss this issue in Appendix A.} slice $\T$ of a parabolic, hyperbolic or elliptic model admitting a symmetry center at $r=0$ is diffeomorphic to the product manifold    
\ba\DD=\eta\times \mathbb{S}^2(\theta,\phi),\qquad
\eta  = \left\{ \begin{array}{l}
 \mathbb{R}^+, \qquad\qquad\quad  \mathbb{R}^3\,\hbox{topology},\\ 
 \left\{ {r\,|\,0 \le r \le r_c } \right\} \quad \mathbb{S}^3\,\hbox{topology}\\ 
 \end{array} \right.\label{DD_def} \ea
Hence, each of these regular $\T$ contains compact concentric spherical comoving regions enclosing a center, bounded by 2--spheres of area $4\pi R^2(t_0,r_0)$, and diffeomorphic to the product manifold    
\ba\DD[r_0]=\eta[r_0]\times \mathbb{S}^2(\theta,\phi)\subset \DD,\nonumber\\
 \eta[r_0] \equiv \{r\,|\, 0\leq r\leq r_0\}\subset \eta,\label{etadef} \ea  
so that every scalar $A$ in $\DD[r_0]$ is equivalent to a real valued function $A:\eta[r_0]\to\mathbb{R}$, and the set $X(\DD[r_0])$ of all scalar functions in $\DD[r_0]$ is equivalent to the set $X(\eta[r_0])$ of all real functions in $\eta[r_0]$. The proper volume of any compact region $\DD[r_0]$ reduces to the following real valued integral on $\eta[r_0]$
\begin{equation} \VV_p(r_0) = \int_{\DD[r_0]}{\dd\VV_p}=4\pi\,\int_0^{r_0}{\FF^{-1}R^2R'\dd x},\label{propvol}\end{equation}
so that $\VV_p(0)=0$.

\begin{quote}

\noindent\underline{Definition 1}. The proper volume average functional associated with a spherical comoving domain $\DD[r_0]$ in a given $\T(t)$ is the linear integral functional $\langle\hskip 0.1 cm\rangle_p: X(\eta[r_0])\to \mathbb{R}$, so that for every scalar function $A\in X(\eta[r_0])$ we get the real number 
\begin{equation}  \Aav_p[r_0]=\frac{\int_{\DD[r_0]}{A\,\dd\VV_p}}{\int_{\DD[r_0]}{\dd\VV_p}}=\frac{\int_0^{r_0}{A\,\FF^{-1}\,R^2\,R'\,\dd x}}{\int_0^{r_0}{\FF^{-1}\,R^2\,R'\,\dd x}},\label{ave_def}\end{equation}
\end{quote}

\noindent\underline{Comment.} Notice that $\langle\hskip 0.1 cm\rangle_p$ is a functional, hence it acts in a non--local manner by associating the real number $\Aav_p[r_0]$ to the scalar $A$ for the whole domain $\DD[r_0]$. If we consider the same scalar $A$ but with its domain given by another comoving region, $\DD[r_1]$ with $r_1\ne  r_0$, then we obtain by means of $\langle\hskip 0.1 cm\rangle_p$ the real number $\Aav_p[r_1]\ne \Aav_p[r_0]$. For all domains (\ref{etadef}) the number $r_0$ marks the 2-sphere (fiber) that is the boundary of a region $\DD[r_0]$. Since $r_0$ is a fixed but arbitrary parameter, if it varies then we can always construct from the average functional a local valued function of $r_0$.\\

\begin{quote}

\noindent\underline{Definition 2}. For every scalar $A\in X(\eta[r])$, the proper average function (``p--function'') is the real valued function $A_p: X(\eta[r])\to \mathbb{R}$, so that for every domain $\eta[r]$  
\begin{equation} A_p(r)=\Aav[r],\label{Apdef}\end{equation}
where $r\geq 0$ encompasses all the domain of regularity of the radial coordinate given by (\ref{DD_def}). 

\end{quote}

\noindent\underline{Comment.} The functional $\Aav[r_0]$ and the function $A_p(r)$ are closely related, though there are subtle but important differences between them. Understanding their similarities and differences is crucial for the proper understanding of the present article, hence we illustrate graphically both objects in figure 1. Notice that for every domain $\eta[r_0]$, we have $A_p(r_0)=\Aav[r_0]$, but $A_p(r)\ne\Aav[r_0]$ for all $r\ne r_0$.\\ 

\noindent\underline{Notation.} In order to avoid confusion and to simplify notation, we will adopt the following conventions: we will use the symbol $\Aav$ without explicit mention of ``$[r_0]$'' (unless it is needed for clarity), as it is evident that $\Aav$ is the real number associated by (\ref{ave_def}) with a domain $\DD[r_0]$ defined by (\ref{etadef}). The functions $A_p$ will be denoted ``p--functions'', reserving the term ``average'' only for the functional $\Aav$. We will use $r$ as independent variable of the functions $A_p$, reserving  $x$ or  for the integration dummy variable.  \\  

\begin{figure}[htbp]
\begin{center}
\includegraphics[width=4in]{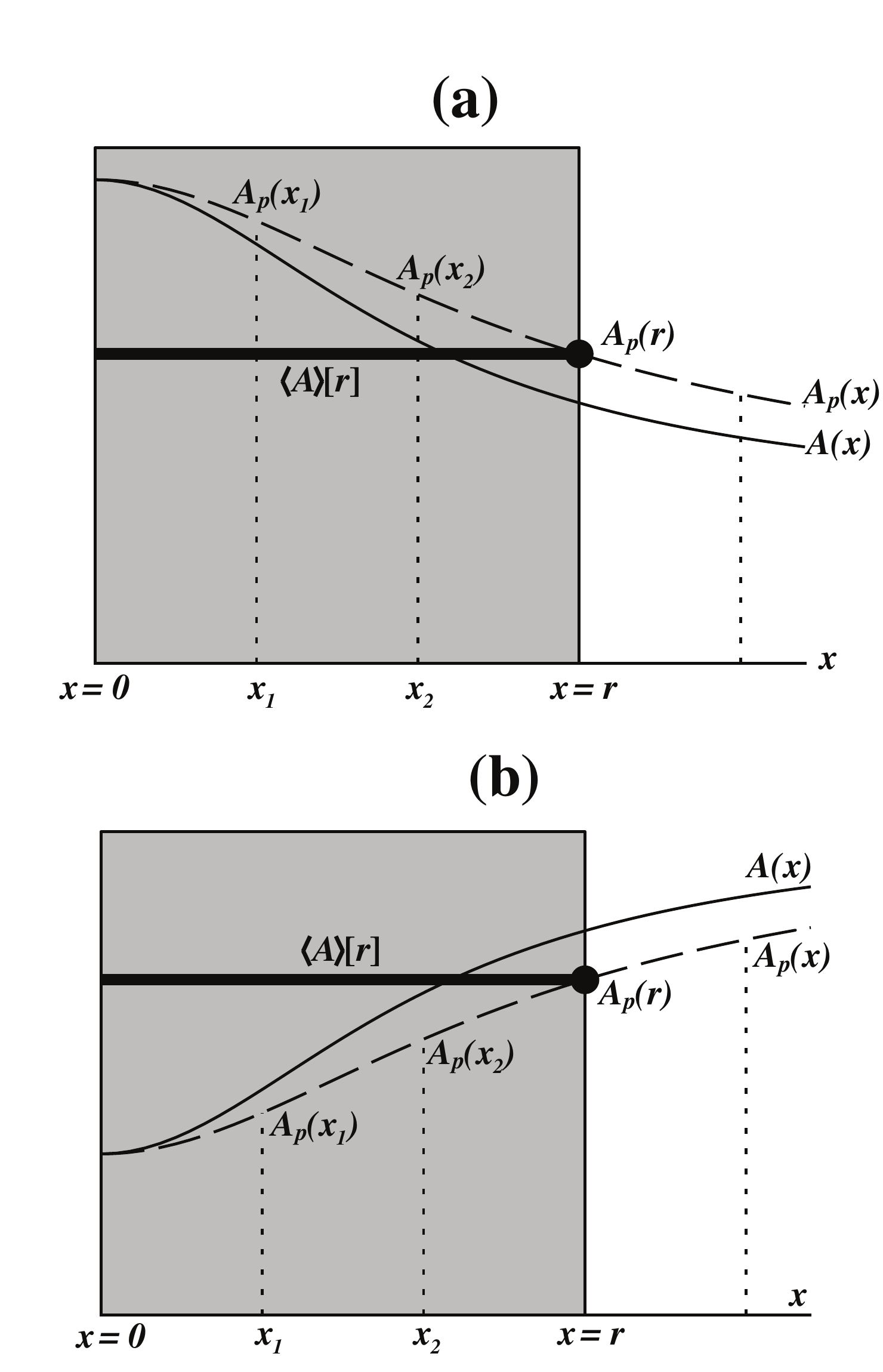}
\caption{{\bf The difference between $A_p$ and $\Aav$.} The figure displays the radial profile of a scalar function $A(x)$ (solid curve) along a regular hypersurface $\T(t)$,  together with its dual auxiliary function $A_p(x)$ (dotted curve) defined by (\ref{Apdef}). Panels (a) and (b) respectively display the cases when $A'\leq 0$ (``clump'') and $A'\geq 0$ (``void''). The average functional (\ref{ave_def}) assigns the real number $\Aav[r]$ to the full domain (shaded area) marked by $\eta[r]=\{x\,|\, 0\leq x\leq r\}$, whereas the function $A_p$ varies along this domain. Hence, $A_p$ and $\Aav$ are only equal at $x=r$, and so they satisfy the same differentiation rules locally, {\it i.e.} $\dot A_p(r)=\Aav\,\dot{}[r]$ and $A_p'(r)=\Aav'[r]$, but behave differently when integrated along the domain. Notice that, as stated in Lemmas 2b and 2c in section 8, and from (\ref{prop2}) and (\ref{prop3}), if $A'\leq 0$ in all $\eta[r]$ then $A-\Aav\leq 0$ and the opposite situation occurs if $A'\geq 0$. This is also true for the quasi--local functions $A_q$.}
\label{fig1}
\end{center}
\end{figure}

Only the functionals $\Aav$ can be considered averages of real valued functions $A$ taken as a continuous random variables. This follows from the fact that $\Aav$ and $A_{{\rm{p}}}$ behave differently under integration along any $\eta[r]$:
\begin{equation} \fl \int_0^{r}{\Aav \,B\, \dd x}=\Aav\int_0^{r}{ B\, \dd x}\quad\hbox{but}\quad \int_0^{r}{A_p \,B \,\dd x}\ne A_p\int_0^{r}{ B\, \dd x}.\end{equation}
Hence, $\Aav$ defined as functionals by (\ref{ave_def}) allow for the construction of momenta such as the variance and covariance, as they comply with
\bse\ba \langle\Aav\rangle=\Aav,\label{avav}\\
\langle \left(A-\Aav\right)^2\rangle=\langle A^2\rangle-\Aav^2,\label{var}\\
 \langle \left(A-\Aav\right)\left(B-\Bav\right)\rangle=\langle AB\rangle-\Aav\Bav, \label{covar}\ea\ese
On the other hand, the functions $ A_{{\rm{p}}}(r)$ do not satisfy (\ref{avav})--(\ref{covar}), and cannot be ``averages'' of a continuous random variable. The proper volume (\ref{propvol}) satisfies the following properties:
\bse\ba \frac{\dot\VV_p}{\VV_p}=\Thetaav,\label{Vdot}\\
\frac{\VV_p'}{\VV_p}=\frac{3R'}{R}\,\frac{\FFav}{\FF},\label{Vprime}\ea\ese  
where (\ref{Vdot}) follows from $\Theta=[\ln(\FF^{-1}R^2R')]\,\dot{}$ and the commutation of $\partial/\partial t$ with the integrals in (\ref{ave_def}), while (\ref{Vprime}) is readily obtained from $\Fav=4\pi R^3/(3\VV_p)$. The commutation rule between the time derivative and the average is
\begin{equation}\Aav\,\dot{} - \langle\dot A\rangle=\langle\Theta A\rangle-\Thetaav \Aav,\label{tconm}\end{equation}
which follows directly by applying $\partial/\partial t$ into (\ref{ave_def}). The functional $\Aav$ and the functions $A_p$ comply with the following properties
\bse\ba    
 \Aav{}' = \frac{\VV_p'}{\VV_p}\,\left[A-\Aav\,\right],\label{prop2}\\
A(r) - \Aav[r] = \frac{1}{\VV_p(r)}\int_0^r{A' \,\VV_p \,\dd x},\label{prop3}\ea\ese 
which follow directly by applying $\partial/\partial r$ and integrating (\ref{ave_def}) by parts.

\section{Quasi--local average functionals and functions.} 

The function $M$ that appears in (\ref{fieldeq1})--(\ref{fieldeq2}) is for LTB models the Misner--Sharp quasi--local mass--energy function, which is an important invariant in spherically symmetric spacetimes. It is basically the volume integral of the field equation (\ref{fieldeq2}), which will be well defined if we assume the existence of a symmetry center (at $r=0$) and can be given as a proper volume integral ``weighed'' by $\FF$:
\begin{equation} 2M =\frac{2G}{c^2}\int_\DD{\rho\,\FF\,\dd\VV_p}=\kappa\int_0^r{\rho R^2 R'\dd x}.\label{MSmass}\end{equation}
This integral definition motivates the introduction of a ``weighed'' average functional and its associated local function. These functions will be very useful in the following sections.\\

\begin{quote}

\noindent\underline{Definition 3}. The quasi--local proper volume average functional associated with a spherical comoving domain $\DD[r_0]$ is the linear integral functional $\langle\hskip 0.1 cm\rangle_q: X(\eta[r_0])\to \mathbb{R}$, so that for every scalar $A\in X(\eta[r_0])$ we get the real number 
\begin{equation} \Aav_q[r_0]=\frac{\int_0^{r_0}{A\,\FF\,\dd\VV_p}}{\int_0^{r_0}{\FF\,\dd\VV_p}}=\frac{\int_0^{r_0}{A R^2R'\,\dd r}}{\int_0^{r_0}{R^2R'\dd r}},\label{aveq_def}\end{equation}

\noindent\underline{Definition 4}. For every scalar $A\in X(\DD[r])$, the quasi--local average function (``q--function'') is the real valued function $A_q: X(\eta[r_0])\to \mathbb{R}$, so that for every $r\in\eta[r]$ 
\begin{equation} A_q(r)=\Aav_q[r]\end{equation}
As with the p--functions, the above definition holds for all $\eta[r]$ and their domain of definition is the full the domain of regularity of $r$ in (\ref{DD_def}).\\

\end{quote} 

The quasi--local volume averages and q--functions satisfy analogous properties to $\Aav$ and $A_p$: only the functionals $\Aav_q$ are averages which comply with (\ref{avav})--(\ref{covar}) with $\langle\hskip 0.1 cm\rangle_q$ replacing $\langle\hskip 0.1 cm\rangle$. The comparison between $\Aav$ and $A_p$ in figure 1 applies also to $\Aav_q$ and $A_q$.  The quasi--local volume is given by
\begin{equation} \VV_q(r) = \int_{\DD[r]}{\FF\,\dd\VV_p}=4\pi\int_0^r{R^2R'\dd x}=\frac{4\pi}{3}R^3(r),\end{equation}
and its derivatives are
\bse\ba \frac{\dot\VV_q}{\VV_q}=\frac{3\dot R}{R}=\Theta_q,\label{Vqdot}\\
\frac{\VV_q'}{\VV_q}=\frac{3 R'}{R},\label{Vqprime}\ea\ese
where (\ref{Vqdot}) follows from $\Theta=[\ln(R^2R')]\,\dot{}$ and the commutation of $\partial/\partial t$ with the integrals in (\ref{ave_def}) (just as with (\ref{Vdot})). The commutation between $\partial/\partial t$ and the quasi--local average (and q--functions) is exactly the same as (\ref{tconm}). The following relations hold
\bse\ba    
 A_q'=(A_q)' = \frac{\VV_q'}{\VV_q}\,\left[A-A_q\,\right],\label{propq2}\\
A(r) - A_q(r) = \frac{1}{\VV_q(r)}\int_0^r{A' \,\VV_q \,\dd x},\label{propq3}\ea\ese 
which are analogous to (\ref{prop2})--(\ref{prop3}). 

Applying the definition (\ref{aveq_def}) to (\ref{fieldeq1}), (\ref{3Ricci1}) and (\ref{MSmass}) we obtain the following important relations
\bse\ba \frac{\kappa}{3}\,\rho_q =\frac{2M}{R^3},\label{qrho}\\
\RR_q = -\frac{6[\FF^2-1]}{R^2}\label{qRR}\\
\frac{1}{9}\Theta_q^2=\frac{\dot R^2}{R^2}=\frac{\kappa}{3}\,\rho_q-\frac{1}{6}\,\RR_q=\frac{2M}{R^3}+\frac{\FF^2-1}{R^2}.\label{Hamcon}\ea\ese
The scalars $\Sigma$ and $\EE$ in (\ref{Sigma1}) and (\ref{EE1}), associated with the shear and electric Weyl tensors, become expressible as deviations of the local scalars $\Theta,\rho$ with respect to their quasi--local duals:
\bse\ba  \Sigma = -\frac{1}{3}\left(\Theta-\Theta_q\right),\label{Sigma2}\\
\EE = -\frac{\kappa}{6}\left(\rho-\rho_q\right).\label{EE2}\ea\ese
These relations among quasi--local q--functions will be very useful in the forthcoming sections.

\section{Buchert's averaging formalism.} 

The idea behind Buchert's spatial averaging formalism is to apply the averager proper volume functional (\ref{ave_def}) to scalar evolution equations for covariant scalars in spacetimes under a suitable time splitting (see \cite{buchert,ave_review}).   For the case of LTB models under consideration, and given the time slicing furnished by $u^a$ (see section 3), the functional (\ref{ave_def}) must be applied to both sides of the scalar evolution equations: the Raychaudhuri and energy balance equations, and to the Hamiltonian constraint (see equations (\ref{ev_theta_13}), (\ref{ev_mu_13}) and (\ref{cHam_13}) in Appendix C). Using the time derivation rule (\ref{tconm}) and (\ref{avav})--(\ref{covar}), we obtain  the evolution laws for $\Thetaav$ and $\rhoav$ and the averaged Hamiltonian constraint~\footnote{These equations follow from the fulfillment of the average properties (\ref{avav})--(\ref{covar}), therefore they do not involve the p--functions $\Theta_p,\,\rho_p,\,\RR_p$. } 
\ba \Thetaav\,\dot{}+ \frac{\Thetaav^2}{3}=-\frac{\kappa}{2}\left[\rho_{\rm{eff}}+3\,P_{\rm{eff}}\right],\label{ave_eveqs_ef1}\\
\rhoav\,\dot{} = -\rhoav\,\Thetaav,\label{ave_eveqs_ef2}\\
 \frac{\Thetaav^2}{9} = \frac{\kappa}{3}\,\rho_{\rm{eff}},\label{ave_eveqs_ef3}\ea     
where the ``effective'' density and pressure are
\ba \kappa\,\rho_{\rm{eff}} &\equiv& \kappa\,\rhoav -\frac{\RRav+\QQ}{2},\label{rhoefe}\\
\kappa\,P_{\rm{eff}} &\equiv& \frac{\RRav}{6}-\frac{\QQ}{2},\label{Pefe}\ea     
and the kinematic ``back--reaction'' term, $\QQ$, is given by
\begin{equation} \QQ \equiv \frac{2}{3} \langle \left(\Theta-\Thetaav\right)^2\rangle -6\langle\Sigma^2\rangle,\label{QQgen}\end{equation}
where $6\Sigma^2=\sigma_{ab}\sigma^{ab}$ follows from (\ref{Sigma1}). The integrability condition between  (\ref{ave_eveqs_ef1}), (\ref{ave_eveqs_ef3}), (\ref{rhoefe}), (\ref{Pefe}) and (\ref{QQgen}) is the following relation between $\dot\QQ$ and $\RRav\,\dot{}$
\begin{equation} \dot\QQ +2\Thetaav\,\QQ+\frac{2}{3}\Thetaav\,\RRav+\RRav\,\dot{}=0,\label{consistQR}\end{equation}
which can be proven to be compatible with the fluid flow evolution equations given in Appendix C.  

Equations (\ref{ave_eveqs_ef1}) clearly convey the motivation of Buchert's approach: by averaging the scalar evolution equations (see Appendix C) we get for a simple dust source averaged quantities, $\Thetaav$ and $\rhoav$, whose evolution mimics that of a source in which there is ``effective'' density and pressure, (\ref{rhoefe}) and (\ref{Pefe}), constructed from ``back--reaction'' terms that arise in the same averaging process. These terms could yield, in principle, a form of positive acceleration in (\ref{ave_eveqs_ef1}) that arises simply from the averaging of inhomogeneities and not from an elusive source like dark energy.   

From (\ref{rhoefe}) and (\ref{Pefe}), the condition for an ``effective'' cosmic acceleration that could mimic dark energy is 
\begin{equation}A_{\rm{eff}}\equiv -\frac{\kappa}{2}\,\left[\rho_{\rm{eff}}+3\,P_{\rm{eff}}\right] > 0,\qquad\Rightarrow\qquad \QQ -\frac{\kappa}{2}\,\rhoav\,>\,0,\label{efeacc}\end{equation}
so that, given a domain of the type (\ref{etadef}) and $\rho\geq 0$ holding everywhere, the necessary (but not sufficient) condition for (\ref{efeacc}) is
\begin{equation} \QQ \geq 0. \label{QQpos}\end{equation}
Therefore, once we find conditions (necessary or sufficient or both) for the fulfillment of (\ref{QQpos}), these will be necessary for (\ref{efeacc}). We examine these conditions for the remaining of the article.

It is important to notice that both $\QQ$ and $A_{\rm{eff}}$ in (\ref{efeacc})  and (\ref{QQpos}) can be expressed as sign conditions on averages of a single scalar defined in a domain $\eta[r]$ of the form (\ref{etadef})
\ba \QQ &=& \frac{2}{3}\left\langle \C_*\right\rangle \geq 0,\label{QQpos2}\\
 A_{\rm{eff}} &=& \langle \ACal_*\rangle > 0\label{efeacc2},\ea
with the scalars $\C_*$ and $\ACal$ given by
\ba \C_* &=& \C_*(x,r) \equiv \left(\Theta(x) -\Thetaav[r]\right)^2- \left(\Theta(x) -\Theta_q(x)\right)^2,\label{Cdef}\\
\ACal_*  &=& \ACal_*(x,r) \equiv   \frac{2}{3}\C_*(x,r) - \frac{\kappa}{2}\rho(x).\label{Edef}\ea 
where we used (\ref{Sigma2}) to express the shear scalar $\sigma_{ab}\sigma^{ab}=6\Sigma^2$ in terms of $\Theta$ and $\Theta_q$.     

\section{Sufficient conditions for a non--negative back--reaction.}

Conditions (\ref{QQpos2}) and (\ref{efeacc2}) have been examined on LTB models by means of approximations \cite{LTBave2} or particular exact solutions \cite{LTBave1,LTBave3}. For a more general theoretical framework, it is practically evident that finding the general (necessary and sufficient) conditions for the fulfillment of these conditions cannot be done without resorting to numerical methods, as it requires  evaluating average integrals (\ref{ave_def}) for fully general metric functions, like $R$ and $R'$, which are known (at best) in implicit or parametric form from solving (\ref{fieldeq1}).       

However, if we are interested in finding sufficient conditions for  (\ref{efeacc})--(\ref{efeacc2}), it is (fortunately) not necessary to compute integrals like (\ref{ave_def}). If what is needed is simply to find out the sign of an averaged quantity $\Aav$ in a given domain $\eta[r]$, we can obtain sufficient information on this sign simply by looking at the behavior of $A$ point by point in the domain. Concretely, we will use the following property of integrable functions $A$ defined in domains like (\ref{etadef}):
\begin{equation} A(x)\geq 0 \quad \forall\quad x\in \eta[r] \quad \Rightarrow\quad \Aav[r] \geq 0, \label{suf_ave_gen}\end{equation}
though, it is important to mention that the converse is not true: we can have $\Aav[r] \geq 0$ even if $A<0$ holds in subsets of $\eta[r]$. We will use the property (\ref{suf_ave_gen}) to examine sufficient conditions for (\ref{QQpos2}), and then look at their implications for (\ref{efeacc2}). 

Sufficient (but not necessary) conditions for (\ref{QQpos2}) and (\ref{efeacc2}) are given by 
\ba\C_*(x,r) \geq 0,\label{QQpos3}\\
\ACal_*(x,r) \geq 0,\label{eFeacc3}\ea  
holding for all $x\in\eta[r]$. Notice that if we prove (\ref{QQpos3}) and/or (\ref{eFeacc3}), then $\QQ\geq 0$ and/or $A_{\rm{eff}}\geq 0$ follow automatically from (\ref{suf_ave_gen}) as a corollary.  

Testing conditions (\ref{QQpos3})--(\ref{eFeacc3}) can be very difficult because $\C_*(x,r)$ is really a family of functions of $x$ for arbitrary fixed $r$. Fortunately, these conditions can be greatly simplified by means of the following:\\

\begin{quote}

\noindent \underline{Lemma 1}: $\langle \W\rangle=0$ in every domain $\eta[r]$ for $\W=\W(x,r)$ given by
\begin{equation} \W(x,r) = \left[\Theta(x)-\Thetaav[r]\right]^2-\left[\Theta(x)-\Theta_p(x)\right]^2,\label{newC}\end{equation}

\noindent \underline{Proof}.  Expanding (\ref{newC}) and applying (\ref{ave_def}) we obtain with the help of (\ref{var})
\begin{equation} \langle\W\rangle[r]=-\Thetaav[r]^2+\frac{1}{\VV_p(r)}\int_0^r{[2\Theta\Theta_p-\Theta_p^2]\,\VV_p'\,\dd x}.\end{equation}
Inserting $\Theta=\dot\VV_p'/\VV_p'$ and $\Theta_p=\dot\VV_p/\VV_p$ in the integrand above, and bearing in mind that $\Thetaav$ and $\Theta_p$ coincide at the domain boundary $x=r$, leads to the desired result:
\begin{equation} \fl \langle\W\rangle[r]=-\Thetaav[r]^2+\frac{1}{\VV_p(r)}\int_0^r{\left[\frac{\dot\VV_p^2}{\VV_p}\right]^\prime\,\dd x}=-\Thetaav[r]^2+\Theta_p^2(r)=0.\end{equation}
An analogous result follows for the quasi--local average acting on a scalar like $\W$ with $\langle\hskip 0.1 cm\rangle_q$ and $\Theta_q$ instead of $\langle\hskip 0.1 cm\rangle$ and $\Theta_p$.\\

\noindent \underline{Corolary}: for any domain $\eta[r]$ we have $\QQ=\langle\C_*\rangle[r]=\langle\C\rangle[r]$, with $\C=\C(x)$ given by 
\begin{equation} \C = \left[\Theta-\Theta_p\right]^2-\left[\Theta-\Theta_q\right]^2,\label{CCdef}\end{equation}
The proof follows directly from Lemma 1, as $\langle(\Theta-\Thetaav[r])^2\rangle=\langle(\Theta-\Theta_p)^2\rangle$.\\

\end{quote}

\noindent
By using Lemma 1, sufficient conditions for $\QQ\geq 0$ given by (\ref{QQpos2}) can be rewritten now in terms of $\C$, which can be given as a function of $r$ (since $x$ is a dummy variable):  
\ba \C(r) &=&  \left[\Theta-\Theta_p\right]^2-\left[\Theta-\Theta_q\right]^2\nonumber\\
 &=&\left[\Theta_q-\Theta_p\right]\,\left[\Theta-\Theta_q+\Theta-\Theta_p\right]\geq 0 \quad\Rightarrow\quad \QQ\geq 0.\label{CC1} \ea
The behavior (sign) of this quantity must be examined for every domain $\eta[r]$.  Considering (\ref{prop3}) and (\ref{propq3}) applied to $\Theta$ we get
\bse\ba \Theta(r)-\Theta_p(r)=\frac{1}{\VV_p(r)}\,\int_0^r{\Theta'(x)\VV_p(x)\,\dd x},\label{TT1}\\
\Theta(r)-\Theta_q(r)=\frac{1}{\VV_q(r)}\,\int_0^r{\Theta'(x)\VV_q(x)\,\dd x},\label{TT2}\ea\ese
and inserting these expressions into (\ref{CC1}) and rearranging terms, we can express this condition as
\ba \C(r) = \Phi(r)\,\Psi(r) \geq 0\quad\Rightarrow\quad \QQ\geq 0,\label{CC2}\\
\Phi(r) \equiv \int_0^r{\Theta'(x)\,\varphi(x,r)\,\dd x},\label{Phidef}\\
\Psi(r) \equiv \int_0^r{\Theta'(x)\,\psi(x,r)\,\dd x},\label{Psidef}\ea
with $\varphi$ and $\psi$ given by
\ba \varphi(x,r)=\frac{\VV_p(x)}{\VV_p(r)}-\frac{\VV_q(x)}{\VV_q(r)}=\frac{\VV_p(x)}{\VV_p(r)}\left[1-\frac{\FF_p(x)}{\FF_p(r)}\right],\label{phi_def}\\
\psi(x,r)=\frac{\VV_p(x)}{\VV_p(r)}+\frac{\VV_q(x)}{\VV_q(r)}=\frac{\VV_p(x)}{\VV_p(r)}\left[1+\frac{\FF_p(x)}{\FF_p(r)}\right],\label{psi_def}\ea
where $\FF_p$ is the p--function associated to $\FF$, and we have used the relation 
\begin{equation} \frac{\VV_q(x)}{\VV_p(x)}=\FF_p(x),\qquad \frac{\VV_q(r)}{\VV_p(r)}=\FF_p(r)\label{prop_1}\end{equation}
which follows directly from (\ref{ave_def}) and (\ref{aveq_def}). Since we need to compare the values of $\FF_p$ in interior points $x$ with those in the boundary of $\eta[r]$ (for arbitrary $\eta[r]$), the  following expressions will be very handy: 
\bse\ba 
\frac{\FF_p'}{\FF_p}=\frac{\VV_q'}{\VV_q}\,\frac{\FF-\FF_p}{\FF}=\frac{3R'}{R}\,\frac{\FF-\FF_p}{\FF},\label{prop_4}\\
\FF(x) - \FF_p(x) = \frac{1}{\VV_p(x)}\int_0^{x}{\FF'(\bar x) \,\VV_p(\bar x) \,\dd \bar x},\label{prop_5}\ea\ese 
We remark that these properties are valid for every $\T(t)$ and for all $x\in\eta[r]$. Notice that, even if $\dot \FF=0$, the function $\FF_p$ involve $\VV_p$ and $\VV_q$ and so is time dependent. However, the radial profiles of $\FF_p$ in all $\T(t)$ are qualitative analogous (see Appendix B), while the profile of $\FF_p$ is analogous to that of $\FF$ (see Lemmas 2a--2c in the following section). 

The fulfillment of (\ref{QQpos3}) is now equivalent to that of (\ref{CC2}), and it clearly depends on the signs of $\varphi$ and $\psi$ (besides the sign of $\Theta'$) at all points in any domain along arbitrary $\T(t)$. Since, by their definition, $\VV_p(0)=\VV_q(0)=0$ and $\FF(0)=1$, (\ref{psi_def}) and  (\ref{phi_def}) imply
\bse\ba \psi(0,r)=0,\quad \psi(r,r)=2,\label{psi11}\\
\frac{\partial}{\partial x}\psi(x,r) = \frac{R^2(x)\,R'(x)}{\FF(x)}\,\frac{\FF_p(r)+\FF(x)}{\VV_p(r)\,\FF_p(r)},\label{psi12}\ea\ese
\bse\ba \varphi(0,r)=\varphi(r,r)=0, \label{phi11}\\
\frac{\partial}{\partial x}\varphi(x,r) = \frac{R^2(x)\,R'(x)}{\FF(x)}\,\frac{\FF_p(r)-\FF(x)}{\VV_p(r)\,\FF_p(r)},\label{phi12}\ea\ese
which indicates that as long as $\FF$ and $R'$ are non--negative, the signs of $\psi$ and $\partial\psi/\partial x$ are non--negative, and so $\Psi$ basically depends on the sign of $\Theta'$. On the other hand, the signs of  $\varphi$ and $\partial\varphi/\partial x$ are not determined, hence the sign of $\Phi$ requires more examination as it depends on both: the sign of $\Theta'$ and the ratio $\FF_p(x)/\FF_p(r)$ (which will depend as well in some cases on the sign of $\FF'$). In elliptic configurations where the $\T(t)$ have spherical topology, $\FF$ and $R'$ can become negative, and so the sign of $\Psi$ is also indetermined (see section 12). Since the sign of $\C$ depends on the sign of the product $\Phi\Psi$, we will need to obtain the conditions for both terms having the same sign.

\section{Radial profiles and turning values (``TV's'').}

The fulfillment of conditions (\ref{QQpos3}) and (\ref{eFeacc3}) (or (\ref{CC2})) is strongly dependent on the radial profiles of scalars, such as $R,\,\FF,\,\FF_p,\,\Theta,\,\Theta_p,\,\Theta_q$ and the volumes $\VV_p$ and $\VV_q$. We will need to probe these profiles along domains $\eta[r]$ in the space slices $\T$ of various LTB configurations. In particular, it is important to examine the cases in which radial gradients can vanish in a domain $\eta[r]$. 

\subsection{Turning values (TV's).}

\begin{quote}

\noindent \underline{Definition 5}.\,\, We will denote by ``Turning Value of a scalar $A$'' (TV of $A$) a value $x=\rtv\in\eta[r]$ such that $A'(\rtv)=0$ under regular conditions (which excludes shell crossing singularities and surface shells, see Appendix A).  

\end{quote}

\noindent
It is important to remark that the TV's for each of $R,\,\FF$ and $\Theta$ occur under different conditions. In general, a given $\eta[r]$ could exhibit either (or all of) these TV's in different coordinate values of $r$. Profiles and TVs of $r$--dependent functions like $\FF(r)$ and $M(r)$ can be directly associated with initial conditions at a fiducial $\T_i=\T(t_i)$ (see Appendix B), hence they are marked by the same comoving radial coordinate in all $\T(t)$. TV's of $R$ only occur in elliptic models in which the $\T(t)$ have topology $\mathbb{S}^3$, and so will also be marked by the same value of $r$. In this case, as shown by the regularity condition (\ref{RrF}), the TV must coincide with a zero of $\FF$ (see section 12 and Appendix B). On the other hand, the TV«s of other scalars ($\Theta,\,\rho,\,\RR$) are not (necessarily) related to initial conditions, hence their coordinate location will (in general) change from one $\T(t)$ to the other. The exception to this rule occurs when there is a TV of $R$, as in this case the TV is common to all other scalars save $\FF$ (see section 12 and Appendices A, B and E and  for further discussion). 

\subsection{ TV's of $\FF$ and spatial curvature.} 

The quantity $1-\FF^2$ and its gradients are directly related to the spatial curvature $\RR$ in (\ref{3Ricci1}) and can be associated with initial conditions. It is useful to examine its relation to $\RR_q$ in (\ref{qRR}), its quasi--local function obtained from (\ref{aveq_def}). From (\ref{3Ricci1}), (\ref{aveq_def}) and (\ref{qRR}) we obtain  
\bse\ba 1-\FF^2 = \frac{1}{6}\,\RR_q\,R^2=\frac{3}{R}\,\int_0^r{\RR\,R^2\,R'\,\dd x},\label{KTVa}\\
\FF' = -\frac{R\,R'}{4\FF}\,\left[\RR-\frac{\RR_q}{3}\right]=-\frac{R\,R'}{6\FF}\,\left[\RR_q+\frac{\RR_q'}{2R'/R}\right].\label{KTVb}\ea\ese
where we used (\ref{propq2}) and (\ref{propq3}) to eliminate $\RR_q'$ in terms of $R',\,\RR$ and $\RR_q$. Equation (\ref{KTVb}) will be used when discussing elliptic models or domains in which TV's of $\FF$ occur (though notice that $\FF$ and $\FF'$ must also comply with the stringent constraints given by (\ref{FFell1}) and (\ref{FFell2a})--(\ref{FFell2c})). \\

\subsection{ TV's of $\Theta$.}

The condition for such a TV (with $R'>0$) come directly from the constraint (\ref{ccons_13}) (Appendix B), which with the help of (\ref{Sigma2}) can be rewriten as 
\bse\ba \Theta' = \Theta_q\left[\frac{3R'}{R}\Dth(1+\Dth)+\left(\Dth\right)'\right],\label{TVTh1}\\
\Dth\equiv \frac{\Theta-\Theta_q}{\Theta_q}=\frac{\Theta_q'/\Theta_q}{3R'/R}.\label{TVTh2}\ea\ese
Since $\Theta_q$ is related through (\ref{Hamcon}) to $M,\,\FF$ and $R$, it can be expressed also in terms of (second order) radial gradients of these functions. A similar condition can be obtained for the TV of $\rho$ through (\ref{ccons_13}) and (\ref{Sigma2})--(\ref{EE2}). Examples of LTB models with TV's of $\rho$ and $\Theta$ are found in \cite{suss08,mustapha}

\subsection{TV's of $A_p$ and $A_q$.}  

We examine in the following Lemmas how the radial profiles and TV«s of scalars $A_p$ and $A_q$ relate to the radial profiles and TV's of $A$:   
    
\begin{quote}

\noindent \underline{Lemma 2a}. If there is no TV of $A$ in $\eta[r]$ then there are no TV's of $A_p$ nor $A_q$.   

\noindent \underline{Lemma 2b}. If $A'\geq 0$ for all $x\in \eta[r]$, then $A(r)\geq A_p(r)$ and $A(r)\geq A_q(r)$.

\noindent \underline{Lemma 2c}. If $A'\leq 0$ for all $x\in \eta[r]$, then $A(r)\leq A_p(r)$ and $A(r)\leq A_q(r)$.\\

\noindent \underline{Proof.} These results follow directly from (\ref{prop3}) and (\ref{propq3}). The converse  statements are not true (for $\eta[r]$). Notice that these results are valid for integration domains like $\eta[r]$ in (\ref{etadef}), which contain a symmetry center. Figure \ref{fig1} illustrates these Lemmas.\\\\

\noindent \underline{Lemma 3}. If there is a TV of $A$ at $x=\rtv$ and $R'>0$ in $\eta[r]$, then for sufficiently large $r$ there will be a TV of $A_p$ at $x=r_1>\rtv$ and a TV of $A_q$ at $x=r_2>\rtv$, with $A(r_1)=A_p(r_1)$ and $A(r_2)=A_q(r_2)$.\\ 

\noindent \underline{Proof.} Let $A'$ pass from positive to negative at $x=\rtv$. As $x$ reaches $\rtv$ the integral in (\ref{prop3}) is still positive and so $A(\rtv)>A_p(\rtv)$, but for $\rtv<x<r$ the integrand becomes negative, and so the contributions to the integral are increasingly negative. Since $\VV(x)/\VV(r)$ is increasing,  if $r$ is sufficiently large, then a value $x=r_1>\rtv$ is necessarily reached so that the integral in (\ref{prop3}) vanishes (thus $A(r_1)=A_p(r_1)$). From (\ref{prop2}), we have $A'_p(r_1)=0$ and $A'_p<0$ for $x>r_1$. An analogous situation occurs when $A'$ passes from negative to positive. The proof is identical for $A_q$, but using (\ref{propq3}) instead of (\ref{prop3}). If $R'=0$ in $\eta[r]$ (there is a TV of $R$), then we can have $A_p'=0$ and $A_q'=0$ with $A_p\ne A$ and $A_q\ne A$. The result of this Lemma is illustrated in figures \ref{fig4}, \ref{fig7} and \ref{fig8}, where we compare the profiles of $\FF$ and $\FF_p$.  

\end{quote}

\section{Parabolic domains and models.}

From (\ref{parab}): $\FF=1,\,\FF'=0$ for all $\eta[r]$. Hence, from (\ref{3Ricci1}) we have $\RR=0$ at all $x\in\eta[r]$ and so these domains are spatially flat. The topological class is necessarily $\mathbb{R}^3$.
\begin{quote}

{\underline{Proposition 1}}: \,\,$\C= 0$ holds in all regular parabolic domains $\eta[r]$.\\ 

\end{quote}
\noindent
{\underline{Proof.}} 

\noindent
If $\FF=1$ holds for all $x\in\eta[r]$, equations (\ref{prop_4}) and (\ref{prop_5}) imply $\FF_p=1$. Hence, $\varphi(x,r)$ in (\ref{phi_def}) vanishes identically in $\eta[r]$, and so $\Phi$ and $\C$ also vanish. Likewise, if $\FF=1$ for $x\in\eta[r]$, then $A_p=A_q$ and so $\C$ is identically zero.\\ 

\noindent
{\underline{Note.}} 

\noindent
It is important to remark that the converse of Proposition 1 is false: $\C=0$ holding in a given domain $\eta[r]$ does not imply that the domain is parabolic. It is possible to obtain a vanishing $\varphi$ for specific cases with $\FF\ne 1$. This possibility was overlooked in \cite{LTBave2,LTBave3}.

\section{Hyperbolic domains and models.}

From (\ref{hyperb}), hyperbolic domains and models are characterized by $\FF\geq 1$, with $\FF(0)=1$. The topological class is necessarily $\mathbb{R}^3$. From (\ref{RrF}) and also from demanding absence of shell crossings~\cite{ltbstuff1,ltbstuff2,ltbstuff3,ltbstuff,suss02} (see Appendix A) we must necessarily have 
\begin{equation} \FF'\geq 0 \quad \forall\,x\in\eta[r],\label{Fhyp}\end{equation}
with $\FF'=0$ only at $r=0$ (or possibly as $r\to\infty$). Regularity conditions require also absence of TV's of $R$. From (\ref{3Ricci1}) and (\ref{RrF}) we have
\begin{equation} \hbox{hyperbolic and regular domain}\quad \Rightarrow\quad \RR\leq 0\quad \forall\,x\in\eta[r],\end{equation}
but the converse is not true, as $\RR$ can be negative in regions of elliptic models (see section 13 and figure \ref{fig9}). However, from (\ref{qRR}) and (\ref{KTVa})--(\ref{KTVb}), for the quasi--local curvature scalar $\RR_q$ we have  
\begin{equation} \hbox{hyperbolic and regular domain}\quad \Leftrightarrow\quad \RR_q\leq 0\quad \forall\,x\in\eta[r].\end{equation}
Regularity conditions \cite{ltbstuff3,ltbstuff} do not allow for a hyperbolic region (enclosing a center) to be surrounded by parabolic or elliptic ``exteriors''. Thus, as long as regularity conditions hold, the existence of regular hyperbolic domains $\eta[r]$ implies a full regular hyperbolic model, with full asymptotic radial range, in which $\RR\leq 0$,\, $\FF\geq 1$ and $\FF'\geq 0$ hold along all the $\T(t)$. Since regularity conditions prevent TV's of $R$ and $\FF$, the only possible TV is that of $\Theta$. We examine first the the case without TV's and then the case of a TV of $\Theta$ 

\subsection{Hyperbolic domains without TV's.}
 
\begin{quote}

{\underline{Proposition 2}}:\,\, $\C\geq 0$ holds in all regular hyperbolic domains $\eta[r]$ without a TV of $\Theta$. 
\end{quote}

\noindent
{\underline{Proof}}. 

\noindent
From (\ref{Fhyp}) and Lemma 2a (previous section): 
\begin{equation} \FF_p'\geq 0 \quad \forall\,x\in\eta[r],\label{Fphyp}\end{equation}
hence for every $\eta[r]$ we have $\FF_p(x)\leq \FF_p(r)$, which together with (\ref{prop_4})--(\ref{prop_5}) and (\ref{psi11})--(\ref{phi12}) implies that $\varphi$ and $\psi$ in (\ref{phi_def}) and (\ref{psi_def}) are both non--negative (see top panel of figure \ref{fig2}). Then, if $\Theta'\geq 0$, we have $\Phi\geq 0$ and $\Psi\geq 0$, and their product is non--negative. If $\Theta'\leq 0$, then $\Phi\leq 0$ and $\Psi\leq 0$ and their product $\C$ is non--negative.\\

\noindent
{\underline{Note}}.  

\noindent
Proposition 2 holds for all domains $\eta[r]$. In this case, $\C\geq 0$ (and consequently) $\QQ\geq 0$ hold for all regular hyperbolic models with monotonical profile of $\Theta$.\\

\subsection{Hyperbolic domains with a TV of $\Theta$.}

\begin{figure}[htbp]
\begin{center}
\includegraphics[width=2.5in]{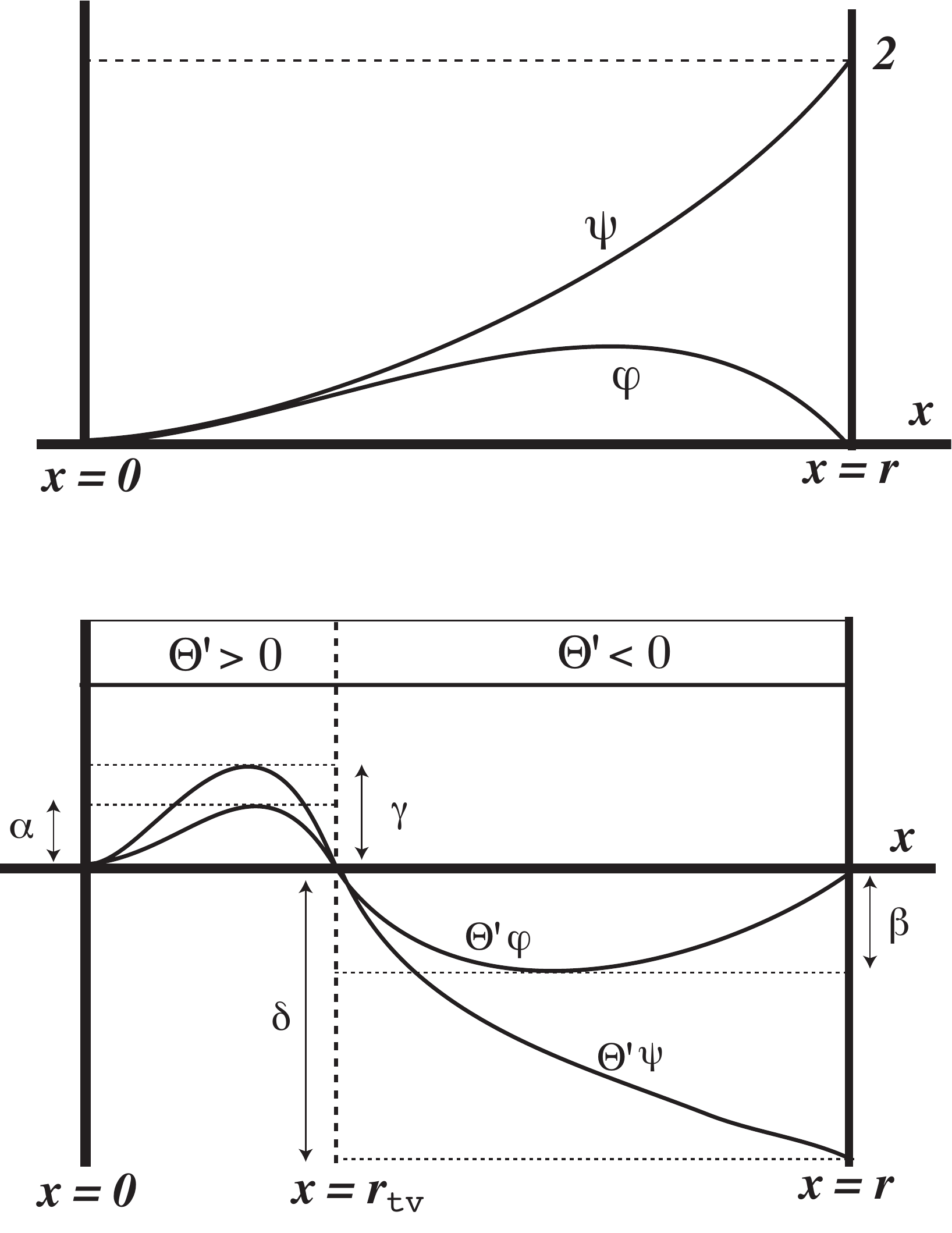}
\caption{{\bf Functions $\varphi$ and $\psi$ for hyperbolic domains.} As shown by the top panel, both functions $\varphi(x,r)$ and $\varphi(x,r)$ are non--negative, hence the signs of $\Phi$ and $\Psi$ depend on the sign of $\Theta'$. The lower panel displays the functions  $\Theta'\,\varphi$ and $\Theta'\,\psi$ (integrands of $\Phi$ and $\Psi$) when there is a TV of $\Theta$ and $\Theta'$ passes from positive to negative at $x=y$ inside $\eta[r]$. For each profile of the integrands we can select four numbers $\{\alpha,\,\beta,\,\gamma,\,\delta\}$ to constrain the integrals. As we show in Proposition 3, it is always possible to find domains $\eta[r]$ for which $\C\geq 0$.}
\label{fig2}
\end{center}
\end{figure}

If there is a TV of $\Theta$ (see (\ref{TVTh1})--(\ref{TVTh2})), then the integrands of $\Phi$ and $\Psi$ no longer have a determined sign. The fulfillment of $\C\geq 0$ becomes domain dependent. Considering (\ref{prop_4})--(\ref{prop_5}) and (\ref{psi11})--(\ref{phi12}), this situation is illustrated by figure \ref{fig2}, whose lower panel shows the profiles of $\Theta'(x)\varphi(x,r)$ and $\Theta'(x)\psi(x,r)$ when $\Theta'$ changes sign in a given $\eta[r]$. We consider this case in the following 
\begin{quote}

{\underline{Proposition 3:}} \,\, In regular hyperbolic models with a TV of $\Theta$ there always exist domains $\eta[r]$, with $0<r\leq r_1$ {\bf{or}} $r\geq r_2$ with $r_1<r_2$, for which $\C\geq 0$ holds.
\end{quote}
\noindent
{\underline{Proof}}. 

\noindent
Let $x=\rtv$ mark the TV of $\Theta$, so that $r/\rtv>1$ (all domains with $r<\rtv$ have no TV's and so are equivalent to those of Proposition 1). We consider the case in which  $\Theta'\geq 0$ for $0\leq x\leq \rtv$ and $\Theta'\leq 0$ for $\rtv\leq x\leq r$ (see bottom panel of figure \ref{fig2}). The case with opposite signs is analogous. Notice that each of the functions $\varphi(x,r)$ and $\psi(x,r)$ in (\ref{phi_def})--(\ref{psi_def}) is really family of functions of $x$ with $r$ appearing as a different fixed parameter for each domain $\eta[r]$. As a consequence, the curves shown in the bottom panel of figure \ref{fig2} (and so the magnitudes of $\{\alpha,\,\beta,\,\gamma,\,\delta\}$) will be different for each domain $\eta[r]$ in any $\T$. However, since these curves must all comply with (\ref{psi11})--(\ref{phi12}) in all domains, they have the same qualitative form as those shown in figure \ref{fig2} in all $\eta[r]$.~\footnote{This qualitative equivalence of the profiles of $\varphi(x,r)$ and $\psi(x,r)$ in all $\eta[r]$ (with or without the existence of TV's) applies to the proofs of all subsequent Propositions.}

For curves as those in the bottom panel of figure \ref{fig2} we can write (\ref{Phidef}) and (\ref{Psidef}) as
\bse\ba \fl \Phi = \int_0^{\rtv}{\Theta'\varphi \,\dd x}+ \int_{\rtv}^r{\Theta'\varphi \,\dd x}=\int_0^{\rtv}{\Theta'\varphi \,\dd x}- \int_{\rtv}^r{|\Theta'|\varphi \,\dd x},\label{split11}\\
\fl \Psi = \int_0^{\rtv}{\Theta'\psi \,\dd x}+ \int_{\rtv}^r{\Theta'\psi \,\dd x}=\int_0^{\rtv}{\Theta'\psi \,\dd x}- \int_{\rtv}^r{|\Theta'|\psi \,\dd x}.\label{split12}\ea\ese
For all such integrals, we can always find four positive real numbers $\{\alpha,\beta,\gamma,\delta\}$ such that (see bottom panel of figure \ref{fig2}):
\bse\ba 0\leq \int_0^{\rtv}{\Theta'\varphi \,\dd x}\leq \alpha\,\rtv,\qquad  0\leq \int_{\rtv}^r{|\Theta'|\varphi \,\dd x}\leq \beta\,(r-\rtv),\label{split21}\\ 
0\leq \int_0^{\rtv}{\Theta'\psi \,\dd x}\leq \gamma\,\rtv,\qquad  0\leq \int_{\rtv}^r{|\Theta'|\psi \,\dd x}\leq \delta\,(r-\rtv). \label{split22}\ea\ese 
Given (\ref{split11})--(\ref{split12}) and (\ref{split21})--(\ref{split22}), it is straightforward to show that $\C=\Phi\Psi\geq 0$ holds for all domains $\eta[r]$ such that 
\begin{equation} \fl \frac{r}{{\rtv}} \leq \rm{min}\left\{\frac{\alpha}{\beta}+1,\,\frac{\gamma}{\delta}+1\right\}\qquad {\hbox{or}}\qquad \frac{r}{{\rtv}} \geq \rm{max}\left\{\frac{\alpha}{\beta}+1,\,\frac{\gamma}{\delta}+1\right\},\label{split31}\end{equation} 
hold, while  $\C=\Phi\Psi\leq 0$ holds for intermediary complementary values of $r/\rtv$. The inequalities (\ref{split31}) provide the required values of $r_1$ and $r_2$ for the domains for any given profile of $\Theta'\varphi$ and $\Theta'\psi$ for a single TV of $\Theta$.  

Notice that the value of $\rtv$ follows from the fulfillment of $\Theta'=0$ in (\ref{TVTh1}) and is fixed at each $\T$, irrespective of the domain $\eta[r]$ that we might choose. This means that in domains with $r>\rtv$, but $r$ close to $\rtv$, the contribution of the areas corresponding to $0\leq x<\rtv$ will dominate the contribution from $\rtv<x<r$ (which accounts for inner zone of the fulfillment of $\C\geq 0$). Since we can choose a domain as extended as we wish ({\it{i.e.}} $r$ can be as large as we want), the curves of $\Theta'\varphi$ and $\Theta'\psi$ for each increasing domain will be like those in figure \ref{fig2}, but with the ratio $\rtv/r$ becoming smaller as $r$ grows. Since there is an asymptotic range of $r$, then for a sufficiently large $r$ the contribution of the area corresponding to $\rtv<x\leq r$ will easily overtake that corresponding to $0\leq x<\rtv$ (which proves the fulfillment of $\C\geq 0$ in the outer asymptotic area). Hence, domains always exist so that $\C>0$ holds in in the inner and outer ranges given by the Proposition.
\\ 

\noindent
{\underline{Note}}.  

\noindent
The most important implication of Proposition 3 is the fact that a TV of $\Theta$ merely leads to the existence of an ``intermediary'' zone with negative $\C$, but no matter how large can this zone be, we can always have $\C>0$ in the appropriate ranges.  If $\Theta$ has several TV's, then there could be a complicated pattern of intermediate zones, but $\C>0$ will always hold (at least) in the asymptotic range as long as there is a clear asymptotic sign of $\Theta'$.

\section{Elliptic domains and models with topology $\mathbb{R}^3$.}  

The regularity condition (\ref{RrF}) implies absence of TV's of $R$, so that we have a full asymptotical range of $r$. The range of $\FF$ in (\ref{elliptic}) reduces to $0\leq \FF\leq 1$, with $\FF(0)=1$ and $\FF=0$ only (possibly) as $r\to\infty$.  There are either no TV's, or TV's of $\Theta$ or $\FF$ or both (only restricted by regularity conditions). Spatial curvature is positive everywhere as long as there are no TV's of $\FF$. We examine first the case without TV's and then cases with TV's.

\subsection{Domains without TV's.} 

If there are no TV's of $\FF$ and $\Theta$, then
\begin{equation} \FF'\leq 0\quad \forall x\in\eta[r],\label{Felp}\end{equation}
with $\FF'=0$ only at $r=0$ (or possibly as $r\to\infty$). Equations (\ref{KTVa})--(\ref{KTVb}) imply that 
\begin{equation} \RR > \frac{\RR_q}{3}=\frac{2\,(1-\FF^2)}{R^2},\label{Fmon}\end{equation}
holds everywhere in $\eta[r]$. Elliptic domains without TV can arise in the following situations:
\begin{itemize}
\item  Near a symmetry center (see Appendix A). Condition (\ref{Fmon}) will hold for all elliptic domains $\eta[r]$ with $r\approx 0$, since $\RR\approx\RR_q\approx \RR(0)$,\, $\Theta'\approx 0$ and $\FF'< 0$.
\item Curvature voids. From (\ref{KTVa})--(\ref{KTVb}) and Lemma 2b, condition (\ref{Fmon}) holds if we have $\RR'>0$ for all $x\in \eta[r]$. 
\item Near homogeneous curvature. If curvature gradients $\RR',\,\RR_q'$ are not large enough, then equations (\ref{propq2})--(\ref{propq3}) imply $\RR\approx\RR_*$ and condition (\ref{Fmon}) holds.  
\end{itemize}
\noindent
However, even in arbitrary elliptic regions or models with turning values of $\FF,\,\Theta$ and $R$, marked by (say) $r_1,\,r_2,\,r_3$, all domains $\eta[r]$ with $r\leq {\rm{min}}(r_1,\,r_2,\,r_3)$ will be domains without TV's. 

\begin{quote}

{\underline{Proposition 4}}:\,\, $\C\leq 0$ holds in all regular elliptic domains $\eta[r]$ without a TV.\\
\end{quote}

\noindent
{\underline{Proof}}. 

\noindent
From (\ref{Felp}) and Lemma 2a, we have 
\begin{equation}  0<\FF_p\leq 1,\quad \FF_p'\leq 0.\label{ell1_reg}\end{equation}  
Therefore, for every $\eta[r]$ we have $\FF_p(x)\geq \FF_p(r)$, which together with (\ref{prop_4})--(\ref{prop_5}) and (\ref{psi11})--(\ref{phi12}) implies that $\varphi\leq 0$ and $\psi\geq 0$. Then, if $\Theta'\geq 0$, we have $\Phi\leq 0$ and $\Psi\geq 0$, and their product is non--positive. If $\Theta'\leq 0$, then $\Phi\geq 0$ and $\Psi\leq 0$ and their product $\C$ is non--positive.\\

\noindent
{\underline{Note}}.  

\noindent
Proposition 4 shows that elliptic domains without TV's and complying with (\ref{Fmon}) have negative back--reaction and so they necessarily exhibit an effective deceleration. Provided the regularity constraint (\ref{FFell1}) (Appendix A) is satisfied, this result becomes domain independent in a given $\T(t)$ if there are no TV's for all domains $\eta[r]$ in it. However, it is extremely likely that in any regular elliptic model the $\T(t)$ will be free from TV«s only for a reduced range of $t$. 
\begin{figure}[htbp]
\begin{center}
\includegraphics[width=2in]{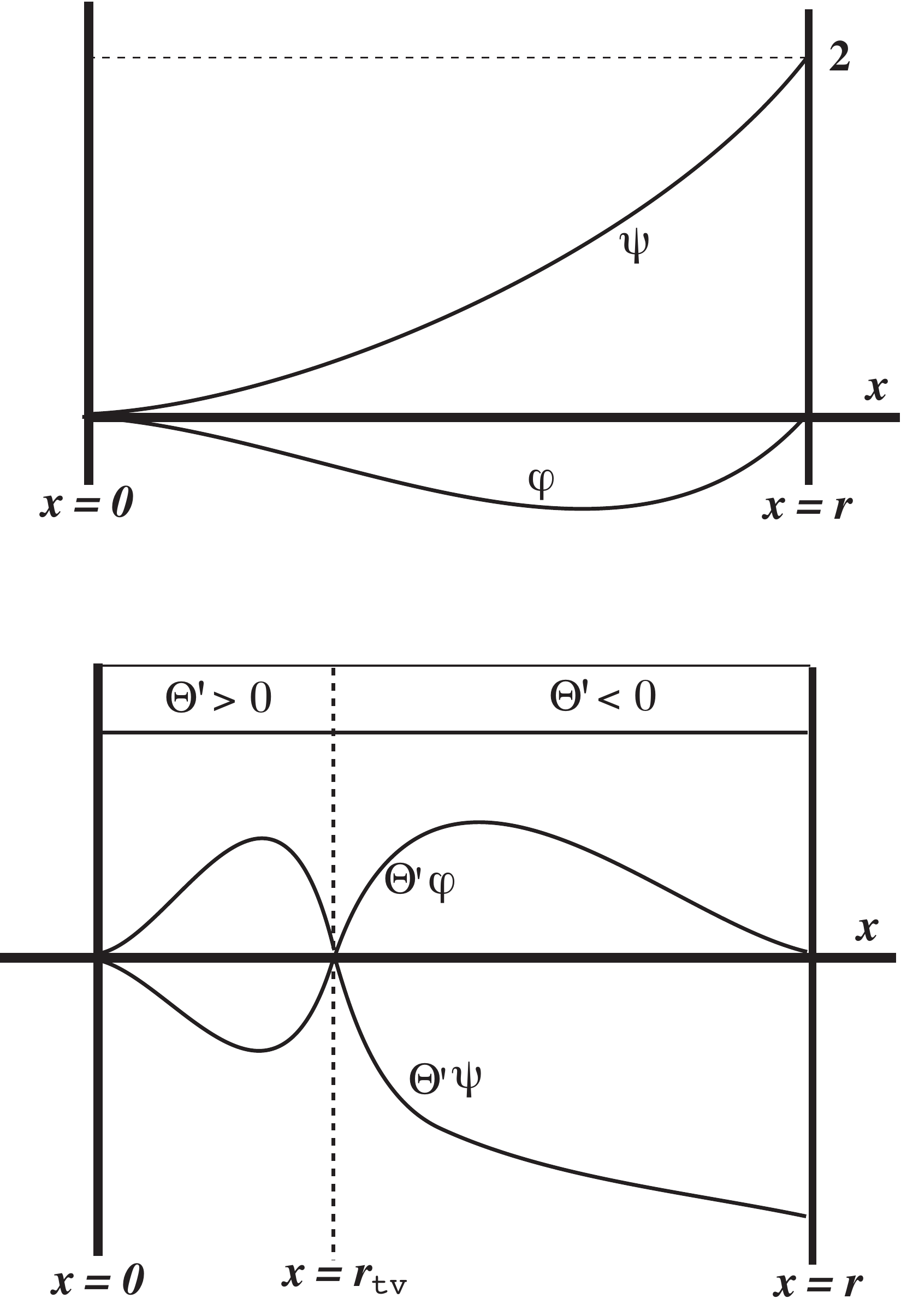}
\caption{{\bf Functions $\varphi$ and $\psi$ for elliptic domains without TV's.} The top panel shows $\varphi(x,r)\leq 0$ and $\varphi(x,r)\geq 0$ for this type of domains. The lower panel displays the functions  $\Theta'\,\varphi$ and $\Theta'\,\psi$ (integrands of $\Phi$ and $\Psi$) when $\Theta'$ passes from positive to negative at $x=y$ inside $\eta[r]$.}
\label{fig3}
\end{center}
\end{figure} 
\begin{figure}[htbp]
\begin{center}
\includegraphics[width=2.5in]{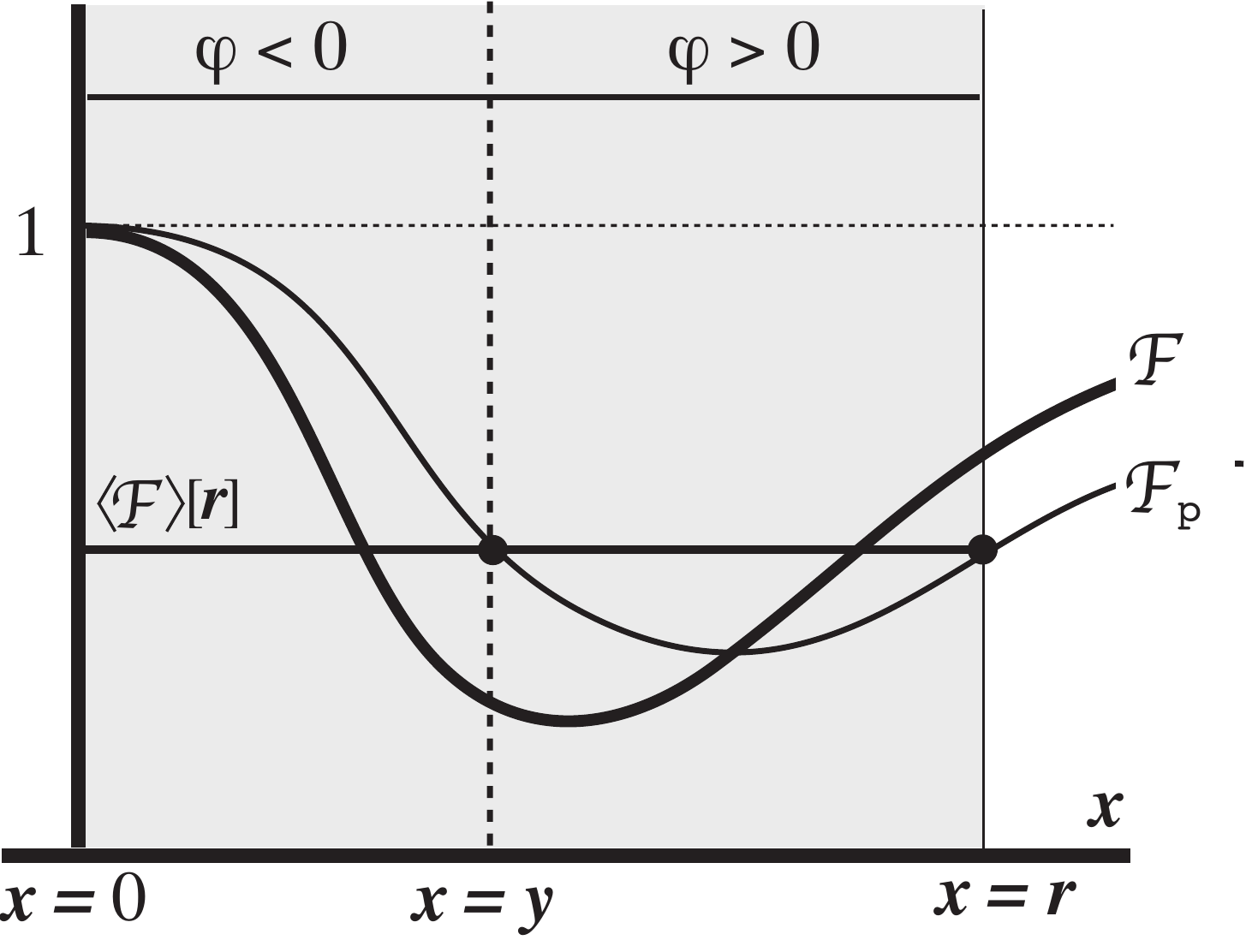}
\caption{{\bf Elliptic domains with a TV of $\FF$.} The figure displays the function $\FF(x)$ (thick curve) when it has a TV, together with its dual function $\FF_p(x)$ (thin solid curve) in a domain $\eta[r]$ (shaded region) in a given hypersurface $\T(t)$. The relation between the profiles and TV's of $\FF(x)$ and $\FF_p(x)$ is given by Lemma 3. Notice that the relation between $\FF_p(x)$ and $\FF_p(r)$ depends on the value $x=y$, as shown in (\ref{ell2_reg}). The location of $y$ is (in general) different from that of the TV's of $\FF$ and $\FF_p$ and it depends on the choice of $r$. From (\ref{phi_def}), the value $x=y$ divides every $\eta[r]$ in two zones: one ($0\leq x<y$) with $\varphi(x,r)\leq 0$ and the other ($y< x\leq r$) with $\varphi(x,r)\geq 0$.}
\label{fig4}
\end{center}
\end{figure}

\subsection{Elliptic domains with a TV of $\Theta$.} 

If there is a TV of $\Theta$ (but no TV's of $\FF$ and $R$), the signs of $\Phi$ and $\Psi$ are no longer determined and the fulfillment of $\C\geq 0$ becomes domain dependent. Considering (\ref{prop_4})--(\ref{prop_5}) and (\ref{psi11})--(\ref{phi12}), this situation is illustrated by the bottom panel of figure \ref{fig3}, displaying the profiles of $\Theta'(x)\varphi(x,r)$ and $\Theta'(x)\psi(x,r)$ when $\Theta'$ changes sign in a given $\eta[r]$. We consider this case in the following  

\begin{quote}

{\underline{Proposition 5}}:\,\, If there is a TV of $\Theta$ in a regular elliptic region at $x=\rtv$, without TV's of $\FF$ or $R$ (open topology), then domains $\eta[r]$ exist for which $\C\geq 0$ holds for $r_1\leq r/\rtv\leq r_2$ with $r_1,\,r_2 >1$. 
\end{quote}
\noindent
{\underline{Proof}}. 

\noindent
We consider the same sign configuration for $\Theta'$ as in Proposition 3. We use (\ref{split11})--(\ref{split12}) and (\ref{split21})--(\ref{split22}) to constrain $\Phi$ and $\Psi$, bearing in mind that now $\varphi\leq 0$ (see figure \ref{fig3}). We readily obtain that $\C=\Phi\Psi\geq 0$ holds for all domains $\eta[r]$ such that 
\begin{equation}   {\rm{min}}\left\{\frac{\alpha}{\beta}+1,\,\frac{\gamma}{\delta}+1\right\} \leq \frac{r}{\rtv} \leq  {\rm{max}}\left\{\frac{\alpha}{\beta}+1,\,\frac{\gamma}{\delta}+1\right\},\label{split41}\end{equation}
Given a profile of $\Theta$ (with one TV) in a domain $\eta[r]$ and its associated ratios $\alpha/\beta$ and $\gamma/\delta$, condition (\ref{split41}) will always hold for elliptic domains in hypersurfaces $\T(t)$ allowing for an asymptotic range (topology $\mathbb{R}^3$), where the ratio $r/\rtv>1$ can be as large or small as required. Notice that Proposition 5 is not affected by the fact that curves in figure \ref{fig3} are different for each $\eta[r]$, as they are qualitatively analogous and $r$ is arbitrary (see comment after Proposition 3). The case with the $\T(t)$ having closed topology is discussed in Proposition 8 and figure \ref{fig6}. \\

\noindent
{\underline{Note}}.  

\noindent
Since elliptic models allow for a collapse of initially expanding dust layers, in general, $\Theta$ for any fixed $r$ will pass from $\infty$ (initial singularity) to $-\infty$ (collapsing singularity), with a likely complex radial dependence pattern along the $\T(t)$ at all times. Thus, even if we assume that there are no TV's of $\FF$ or $R$, it is extremely likely that the $\T(t)$ will exhibit at least a TV of $\Theta$.  Proposition 5 shows that such a TV introduces for some elliptic models with open topology (which satisfy (\ref{Fmon})) an ``intermediary'' range in $r$ with positive back--reaction, while back--reaction remains negative in the region around the center and also in the asymptotic range. As we show below, if there is a TV of $\FF$, then back--reaction can be positive in the full asymptotic region.

\subsection{Elliptic domains with TV's of $\FF$ (but not $\Theta$).} 

We consider first the case when there is only a TV of $\FF$. As shown by (\ref{KTVa})--(\ref{KTVb}), the necessary condition for a TV of $\FF$ is $\RR_*'<0$, which (by virtue of (\ref{propq2})--(\ref{propq3})), implies a negative gradient $\RR'$. If the gradients $\RR'$ and $\RR_*'$ are sufficiently large and negative (assuming regularity), then  $\RR$ and $\RR_*$ will be sufficiently different from each other (from (\ref{propq2})--(\ref{propq3})) to allow for a ``critical value''
\begin{equation} \RR(\rtv) = \frac{\RR_*(\rtv)}{3}, \label{critval}\end{equation}
to occur, where $\rtv$ marks the TV of $\FF$. Hence, we have $\FF'<0$ (or $\RR> \RR_*/3$) for $0<r<\rtv$ and  $\FF'>0$ (or $\RR< \RR_*/3$) for $0<r<\rtv$ (see figure \ref{fig4}). 

While a function $\FF$ with a TV can be prescribed as an initial condition, it is important to remark that $\FF$ and its gradients $\FF'$ cannot be arbitrary, as they are strongly constrained by regularity conditions (see equations (\ref{FFell1}) and (\ref{FFell2a})--(\ref{FFell2c}) in Appendix A). 

\begin{quote}

{\underline{Proposition 6:}} \,\,If there is a TV of $\FF$ in a regular elliptic region, but without a TV of $\Theta$, then domains $\eta[r]$ exist for which $\C\geq 0$ holds for $r\geq a$ for some $a>0$. 
\end{quote}

\noindent
{\underline{Proof}}.

\noindent
Considering Lemma 3 and (\ref{prop_4})--(\ref{prop_5}), the existence of a TV of $\FF$ implies a profile of $\FF_p$ of the form illustrated by figure \ref{fig4}. Bearing in mind (\ref{phi_def}), there always exist $y\in\eta[r]$ (in general $\ne \rtv$) in every $\eta[r]$ and in every $\T(t)$ such that
\ba  \FF_p(x)\geq \FF_p(r),\quad \varphi \leq 0,\qquad 0\leq x\leq y,\nonumber\\ 
 \FF_p(x)\leq \FF_p(r),\quad \varphi \geq 0,\qquad y\leq x\leq r.\nonumber\\\label{ell2_reg}\ea  
The TV of $\FF$ has no effect on the sign of $\psi$, so the profiles of $\varphi$ and  $\psi$ take the forms shown in the top panel of figure \ref{fig5}. The sign of $\C$ then depends only on the sign of $\Phi$, whose integrand  $\varphi$ changes sign due to the presence of the TV of $\FF$ (in agreement with (\ref{ell2_reg})).

 We use (\ref{prop_4})--(\ref{prop_5}), (\ref{psi11})--(\ref{phi12}) and (\ref{ell2_reg}) to set up an analogous construction as in (\ref{split11}) and (\ref{split21}) applied to $\varphi$ and $\psi$ displayed in the top panel of figure \ref{fig5}. Since $\Psi\geq 0$ and there are no TV's of $\Theta$ and $R$, we only need to deal with $\Phi$. Considering the  numbers $\alpha,\,\beta$ to constrain $\Phi$ for the sign change of $\varphi$ at $x=y$ (see figure \ref{fig4}), it is straightforward to show that $\C\geq 0$ holds for all domains $\eta[r]$ such that 
\begin{equation}    \frac{r}{y} \geq  \frac{\alpha}{\beta}+1,\label{split51}\end{equation}
where the ratio $\alpha/\beta$ depends on the profile of $\varphi(x,r)$ and on the value $y$ such that $\varphi(y,r)=0$. As in Propositions 2 and 5, the profiles of $\Theta'\varphi$ and $\Theta'\psi$ (as displayed in \ref{fig4}) will be qualitatively analogous in each domain $\eta[r]$, hence it is always possible to satisfy a condition like (\ref{split51}) because there is a full asymptotic radial range and so $r/y$ can be as large as needed. 

\subsection{Elliptic domains with TV's of $\FF$ and $\Theta$.}

The signs of $\Phi$ and $\Psi$ now also depend on the sign of $\Theta'$ and we have the following

\begin{quote}

{\underline{Proposition 7:}} \,\,If there are TV's of $\FF$ and $\Theta$ in a regular elliptic region, but without a TV of $R$, then domains $\eta[r]$ exist for which $\C\geq 0$ holds for $0<r_1\leq r$ and $r\geq r_2$ with $r_1<r_2$. 
\end{quote}

\noindent
{\underline{Proof}}. 

\noindent
If there is a TV of $\Theta$ marked by $x=\rtv$ (besides the TV of $\FF$), the effect is simply to make $\psi$ change sign once (at the TV of $\Theta$), with an extra sign change in $\Theta'\varphi$. This is illustrated by the lower panel of figure \ref{fig5}. The detail of the demonstration is given in Appendix D, which provides $r_1$ and $r_2$ in terms of $y/\rtv$ and $r/\rtv$ for any given profiles of $\Theta'\varphi$ and $\Theta'\psi$. As in previous Propositions, provided the regularity constraint (\ref{FFell1}) (Appendix A) is satisfied, the fact that the $\T(t)$ have open topology allows for ratios $r/\rtv$ and $y/\rtv$ as large or small as needed to fulfill $\C\geq 0$ (see Appendix D).
\\

\noindent
{\underline{Note}}.  

\noindent
The effect of TV's of $\FF$ and $\Theta$ is to allow for positive back--reaction in the asymptotic range away from the center, in elliptic domains in which $\RR$ shows a sufficiently large decreasing gradient and when hypersurfaces $\T(t)$ have topology $\mathbb{R}^3$. We emphasize again that $\FF$ and $\FF'$ must comply with regularity conditions (see Appendix A).

\begin{figure}[htbp]
\begin{center}
\includegraphics[width=3in]{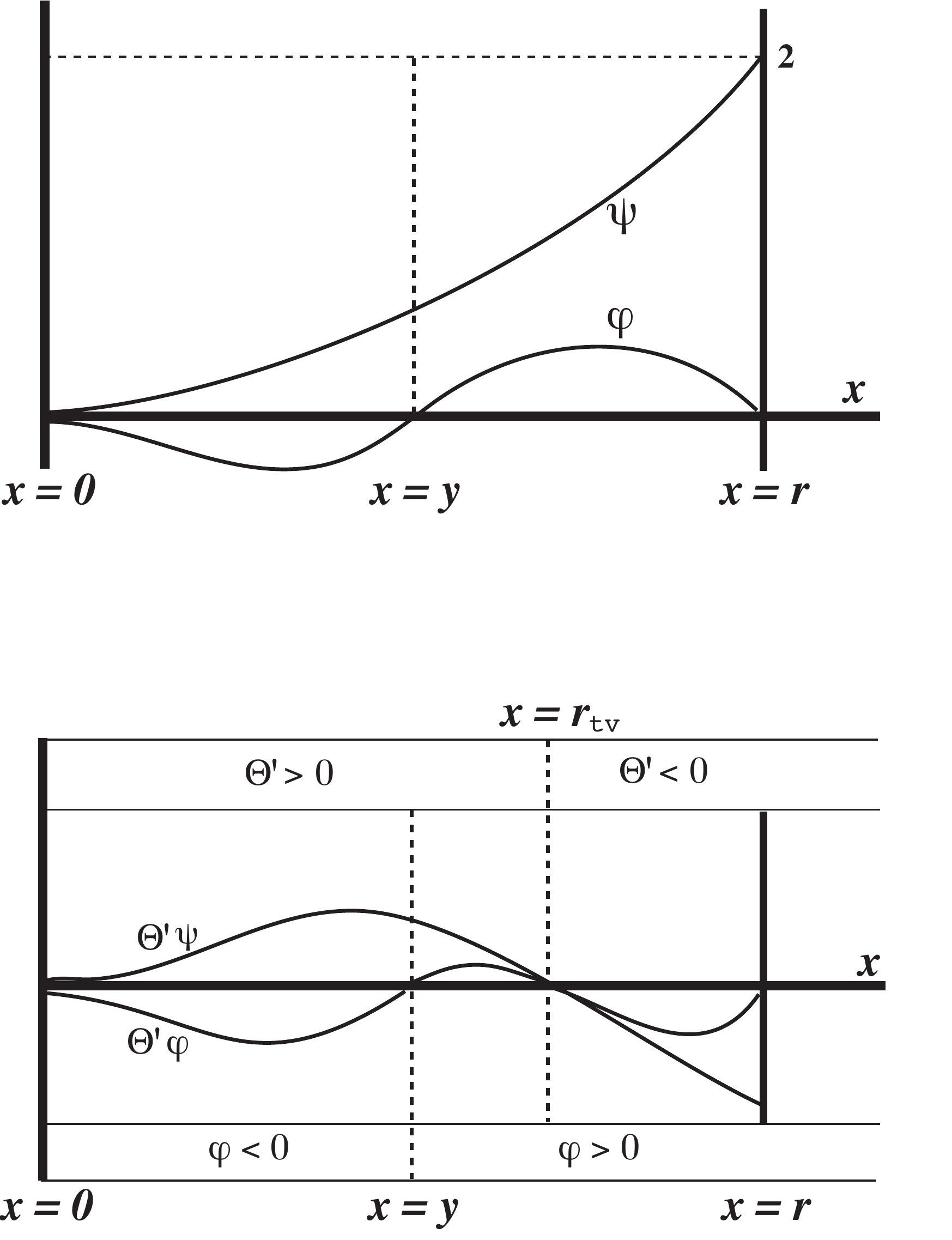}
\caption{{\bf Profiles of integrands of $\Phi$ and $\Psi$ in elliptic domains with TV's of $\FF$ and $\Theta$.} The top panel displays $\phi$ and $\psi$ defined by (\ref{phi_def}) and (\ref{psi_def}) when there is a TV of $\FF$, while the lower panel shows $\Theta'\phi$ and $\Theta'\psi$ when there are TV's of $\FF$ and $\Theta$ (at $\rtv$) with $\Theta'$ passing from positive to negative.}
\label{fig5}
\end{center}
\end{figure}
\begin{figure}[htbp]
\begin{center}
\includegraphics[width=3in]{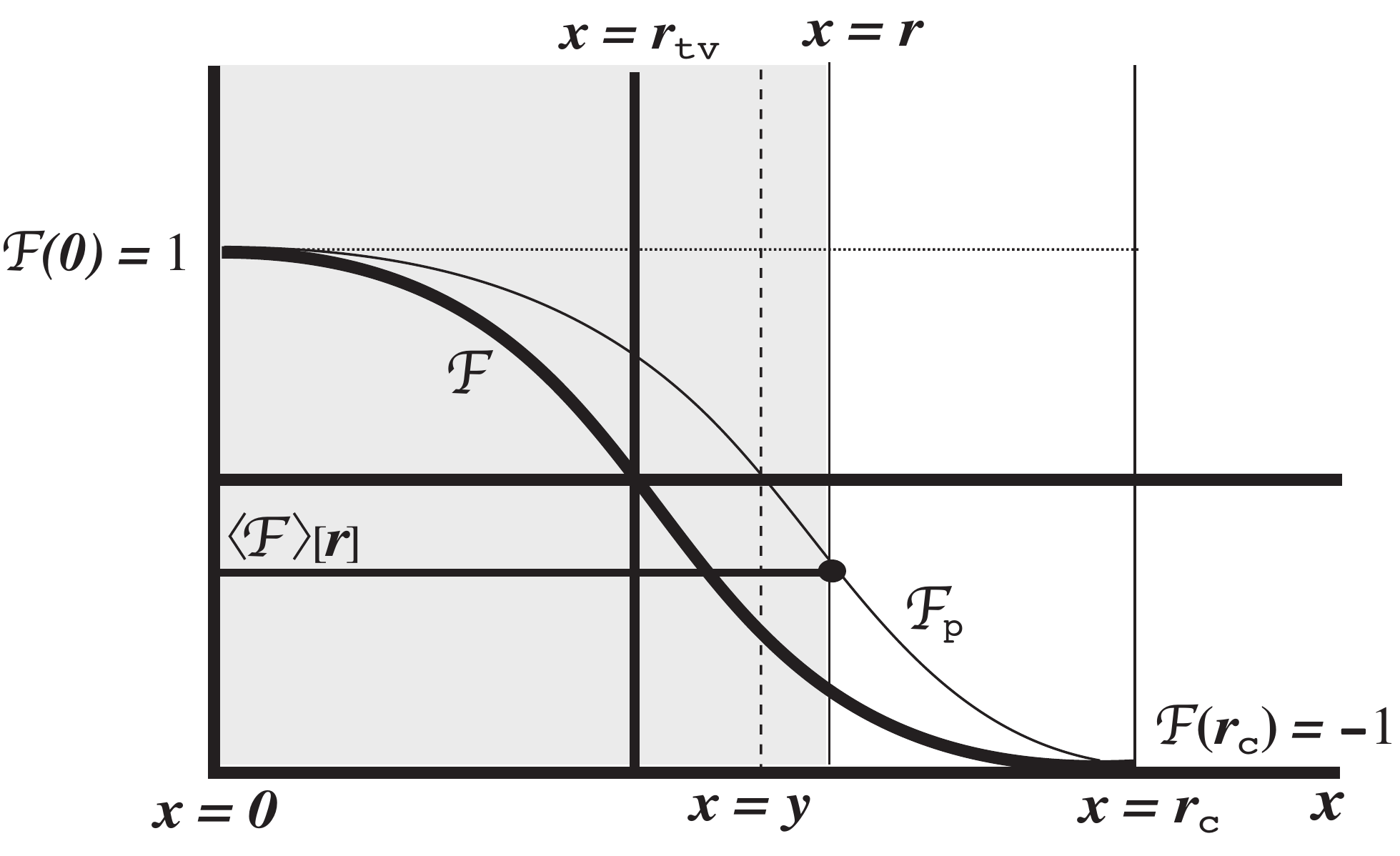}
\caption{{\bf Elliptic domains with a TV of $R$.} The figure displays the function $\FF(x)$ (thick curve) together with its dual function $\FF_p(x)$ (thin solid curve) in a domain $\eta[r]$ (shaded region) with a TV of $R$ marked by $x=\rtv$ (which is then also a zero of $\FF$ and a TV of $\Theta$). Notice that $\FF<0$ for $r>\rtv$ and $\FF_p<0$ for $r>y$. Hypersurfaces $\T(t)$ have $\mathbb{S}^3$ topology with a second symmetry center at $r_c$. Notice that all domains with $r\leq y$ are analogous to those examined in Propositions 4, 5, 6 and 7, whereas in Proposition 8 we considered domains with $r>y$.}
\label{fig6}
\end{center}
\end{figure}

\section{Elliptic models with topology $\mathbb{S}^3$.}

In this case the radial range is restricted by: $0\leq x\leq r_c$, where $x=r_c$ marks the second symmetry center.  Since $R(t,0)=R(t,r_c)=0$, then there must exist a TV of $R$ at $x=\rtv$. The regularity conditions (\ref{RrF}) and (\ref{Rrxi}) necessarily imply that $\FF(\rtv)=0$, and so
\ba R'>0,\quad 0<\FF\leq 1,\qquad\,\, 0\leq x< \rtv,\nonumber\\
R'<0,\quad -1\leq\FF< 0,\qquad \rtv< x\leq \rtv,\nonumber\\\ea 
so that $\FF'\leq 0$ for all $x\in\eta[r]$, with $\FF'(0)=\FF'(r_c)=0$. For elliptic models of this type, this TV of $R$ is also a TV of the scalars $\Theta,\,\rho,\,\RR,\,M$ and all $A_q$, but not $\FF$ and $\FF_p$ (the proof is given in full in Appendix E). However, in locations $r_i\ne \rtv$ there could exist other TV's of $\FF$ (as long as $\FF(\rtv)=0$ holds) or $\Theta$. Depending on the combination of these TV's, some domains $\eta[r]$ can be analogous to the domains of the type examined in Propositions 4, 5, 6 and 7.

\begin{quote}

{\underline{Lemma 4:}} \,\, In any $\T(t)$ of an elliptic model with topology $\mathbb{S}^3$ in which $R'(\rtv)=\FF(\rtv)=0$, there always exist a value $x=y>\rtv$ so that every domain $\eta[r]$ with $r<y$ is analogous to a domain of a $\T(t)$ with topology $\mathbb{R}^3$. We have then the following types of domains for $\eta[r]$ with $r<y$:

\begin{itemize}

\item If $y<\rtv$ and there are no TV's of $\FF$ or $\Theta$ in $\eta[y]$, then $\eta[y]$ is equivalent to an elliptic domain without TV's, as those considered in Proposition 4. In particular, all domains sufficiently close to the center will be of this type.

\item If $y>\rtv$ and there are is no TV's of $\FF$, then there is only the common TV of $R$ and $\Theta$ at $x=\rtv$. In this case, $\eta[y]$ is equivalent to an elliptic domain with a TV of $\Theta$, as those considered in Proposition 5. 

\item If $y<\rtv$ and there is a TV of $\FF$ at $z\ne\rtv$ but no TV of $\Theta$, then $\eta[y]$ is equivalent to an elliptic domain with a TV of $\FF$, as those considered in Proposition 6.

\item If $y>\rtv$ and there is a TV of $\FF$ at $z\in\eta[y]$, then $\eta[y]$ is equivalent to an elliptic domain with TV's of $\FF$ (at $x=z$) and $\Theta$ (at $x=\rtv$), as those considered in Proposition 7.

\end{itemize}

{\underline{Proof:}} \,\, Since $\FF$ is monotonous, then Lemmas 2a and 2c imply
\begin{equation} \FF_p'\leq 0\quad\hbox{and}\quad \FF_p\geq \FF\quad \forall x\in\eta[r]\end{equation}
Thus, there will always exist $y\in\eta[r]$ such that $y>\rtv$ and $\FF_p(y)=0$ hold, so that $\FF_p(x)>0$ for all $x<y$. The profiles of $\FF$ and $\FF_p$ are shown in figure \ref{fig6}. These profiles and the fact that $x=\rtv$ necessarily marks a TV of $\Theta$ imply that the above mentioned combinations of TV's are always possible. \\

{\underline{Note:}} \,\, While Propositions 4--7 apply to all these situations, the restricted radial range $0\leq x\leq y$ implies that conditions of the form (\ref{split41})--(\ref{split51}) or (\ref{AC3})--(\ref{AC4}) could fail to hold (not to mention failure to comply with the regularity constraint (\ref{FFell1}), see Appendix A). Hence, when these domains are part of an elliptic model with closed topology, the wording of Propositions 4--6 must be modified by stating that ``regular domains complying with $\C\geq 0$ {\it{might}} exist''.
  
\end{quote}

If we consider domains $\eta[r]$ with $r>y>\rtv$, where $\FF_p(y)=0$, we have a situation without analogue in elliptic models with open topology because $\FF$ and $\FF_p$ are both negative (see figure \ref{fig6}). Assuming that the only TV is the one marked by $\rtv$ (that is a common TV of $\Theta$), we have the following 

\begin{quote}

{\underline{Proposition 8:}} \,\,If there is a TV of $R$ at $x=\rtv$ in an elliptic model, without other TV's in $r\ne \rtv$ but with $\FF_p(y)=0$ for $y>\rtv$, then domains $\eta[r]$ with $y<r<r_c$ might exist for which $\C\geq 0$ holds. 
\end{quote}

\noindent
{\underline{Proof}}.\,\, $\FF$ and $\FF_p$ monotonously decrease passing from positive to negative (as displayed in figure \ref{fig6}). Since $\FF\leq\FF_p$ but $-1\leq\FF_p(r)<0$, Lemma 4 implies that there is always a value $y>\rtv$ such that $\FF_p(x)<0$ for $y<x<r$. The functions $\varphi(x,r)$ and $\psi(x,r)$ in (\ref{phi_def}) and (\ref{psi_def}) take the forms:
\bse\ba \fl\varphi = \frac{\VV_p(x)}{\VV_p(r)}\left[1+\frac{\FF_p(x)}{|\FF_p(r)|}\right], \quad\psi = \frac{\VV_p(x)}{\VV_p(r)}\left[1-\frac{\FF_p(x)}{|\FF_p(r)|}\right],\quad 0\leq x\leq y,\label{topS31}\\
\fl\varphi = \frac{\VV_p(x)}{\VV_p(r)}\left[1-\frac{|\FF_p(x)|}{|\FF_p(r)|}\right], \quad\psi = \frac{\VV_p(x)}{\VV_p(r)}\left[1+\frac{|\FF_p(x)|}{|\FF_p(r)|}\right],\quad y\leq x\leq r,\label{topS32}\ea\ese
Since $|\FF_p(r)|\leq 1$ and $|\FF_p(x)|\leq |\FF_p(r)|$ in $ y\leq x\leq r$, then $\varphi(x,r)\geq 0$ in all $\eta[r]$. On the other hand, there is also a value $z<y$ such that $\FF_p(x)=|\FF_p(r)|$. Assuming that $z<\rtv$ (the opposite case is analogous), we have $\psi(x,r)\leq 0$ in $0\leq x\leq z$ and $\psi(x,r)\geq 0$ in in $z\leq x\leq r$. Bearing in mind that there is a TV of $\Theta$ at $\rtv$ and assuming that $\Theta'$ passes from postive to negative (the opposite case is analogous), the curves of $\varphi,\,\psi$ and $\Theta'\varphi,\,\Theta'\psi$ are qualitatively analogous to those shown in figure \ref{fig5}. Hence, the proof is the same as that of Proposition 7 with the restriction that $r>\rtv$ (see Appendix D).\\

\noindent
{\underline{Note}}. While the conditions for $\C\geq 0$ are similar to those of domains of elliptic models with topology $\mathbb{R}^3$ examined in Propositions 4--7, these conditions are far more restrictive for the topology $\mathbb{S}^3$. This is so because the parameters $y$ and $z$ change from one $\T(t)$ to the next and are strongly constrained. While we have $r/\rtv>1$, this ratio cannot take arbitrary large values because there is no asymptotic radial range. Since regularity conditions provide extra  constraints on the profiles of $\FF$ and $\FF'$ (see Appendix A), it is practically impossible to make general statements applicable to all profiles and all $\T(t)$. The verification of $\C\geq 0$ for elliptic models with closed topology needs to be done in a case by case basis, though it is likely that regular configurations should exist for which $\QQ\geq 0$ holds, at least for some $\T(t)$.

\section{Elliptic regions in an expanding background.} 

The most interesting configurations that can be constructed with LTB models are perhaps those describing an inner elliptic region possibly undergoing local collapse smoothly immersed in an expanding (parabolic or hyperbolic) background.   

\begin{figure}[htbp]
\begin{center}
\includegraphics[width=3in]{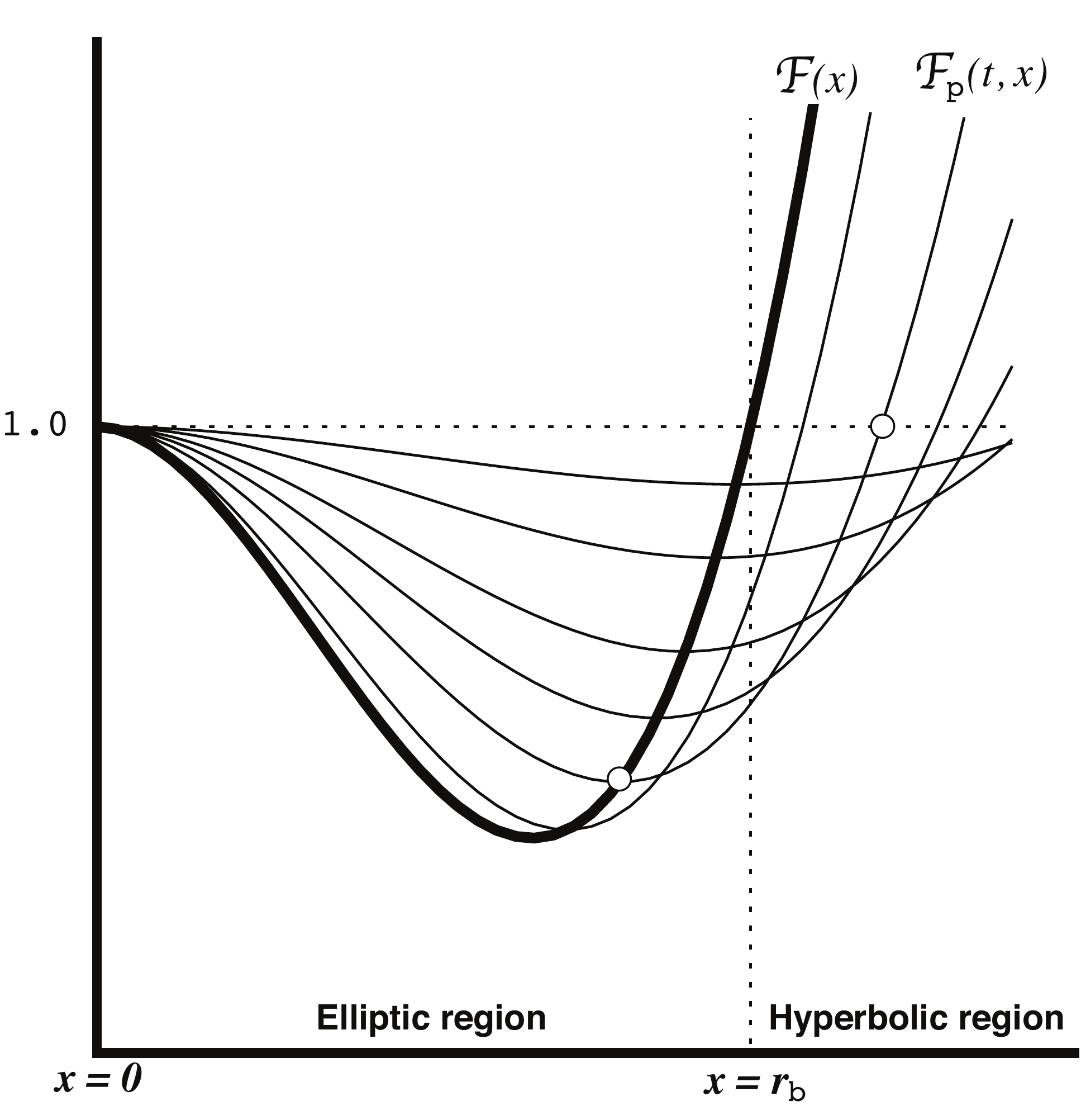}
\caption{{\bf Profiles of $\FF$ and $\FF_p$ in an elliptic region surrounded by a hyperbolic exterior.} The figure displays the radial profile of $\FF$ and those of the time dependent $\FF_p$ for various hypersurfaces $\T(t)$. Since $\FF=\FF(x)$, the profile given by (\ref{Fprof}) is the same for all $\T(t)$. While the profile of $\FF_p$ is different for different $\T(t)$, all share common features that emerge from Lemma 3 and properties (\ref{prop_4}) and (\ref{prop_5}) that are valid for all $\T(t)$. All curves $\FF_p$ intersect $\FF$ when $\FF_p'=0$ in a value $x=y<r_{\rm{b}}$ (white circle to the left), and for all curves there is a value $x=z> r_{\rm{b}}$ for which $\FF_p=1$ (white circle to the right). Different types of domains $\eta[r]$ in a elliptic/hyperbolic configuration are displayed in figure \ref{fig8}. }
\label{fig7}
\end{center}
\end{figure} 
\begin{figure}[htbp]
\begin{center}
\includegraphics[width=4.5in]{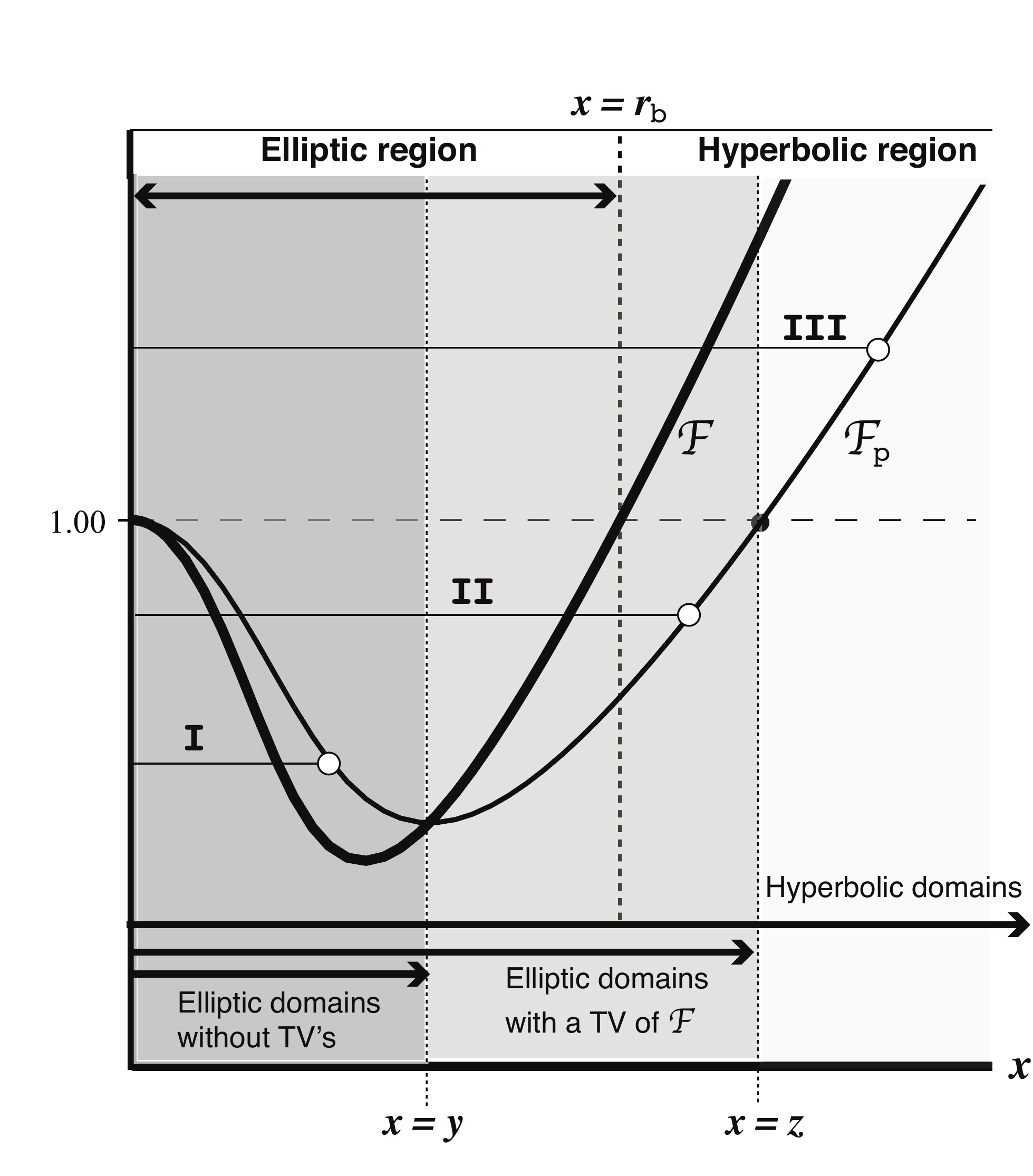}
\caption{{\bf Domains in an elliptic region enclosed by a hyperbolic exterior.} The profile of $\FF(x)$ is shown (thick curve) next to that of a $\FF_p$ in a fiducial $\T(t)$. Assuming absence of TV's of $\Theta$, three types of domains $\eta[r]$ can be constructed, depending on the location of the domain boundary $r$ (white circle). It is evident that domains of the type I comply with (\ref{ell1_reg}) and so are identical to elliptic domains without TV's (Proposition 4), those of the type III behave as hyperbolic domains without TV's (Proposition 2), while domains of type II are identical to elliptic domains with a TV of $F$ (Proposition 6). We can then identify three zones: an internal elliptic one (dark gray) and a hyperbolic external one (very light grey), with an intermediary transition zone. Notice that the boundary of the elliptic region $x=r_{\rm{b}}$ lies in the intermediary zone. If there is a TV of $\Theta$, we have the same situation, but then, depending on the location of the TV, either one of the domains I, II and III would be equivalent to one of those of Propositions 5, 7 and 3.}
\label{fig8}
\end{center}
\end{figure} 

\subsection{Hyperbolic exteriors.}

An elliptic region surrounded by a hyperbolic exterior can be constructed in a single coordinate patch (without ``matching'' as in \cite{ltbstuff3,ltbstuff}) by choosing $\FF$ as 
\begin{equation}
\FF(x) \quad \left\{ \begin{array}{l}
   < 1,\quad 0 \le x < r_{\rm{b}}  \\ 
  = 1,\quad x = r_{\rm{b}}  \\ 
 > 1,\quad x > r_{\rm{b}}  \\ 
 \end{array} \right.\label{Fprof}
\end{equation}
where  $r_{\rm{b}}$ marks the comoving boundary between the regions. We assume that there are no TV's of $\Theta$ and $R$, though $\FF(0)=1$, and so $\FF$ must necessarily have a TV in order to reach $\FF(r_{\rm{b}})=1$ (besides the fact that $\FF$ and $\FF'$ must comply in the elliptic region with regularity constraint (\ref{FFell1}), see Appendix A). We have a similar behavior of $\FF$ and $\FF_p$ as in elliptic domains with a TV of $\FF$, except that now the hyperbolic background allows for domains in which $\FF>1$ for $x > r_{\rm{b}}$. The profiles of $\FF(x)$ and $\FF_p(t,x)$ for various $\T(t)$ are shown in figure \ref{fig7}.  We prove now the following 

\begin{quote}

{\underline{Proposition 9}}:\,\, In all regular LTB configurations made by a hyperbolic region surrounding an elliptic region containing a center, without a TV of $R$ but with possible TV's of $\Theta$, there are always domains $\eta[r]$ with $r>a$ for some $a>0$ for which $\C\geq 0$ holds. 
\end{quote}

\noindent
{\underline{Proof}}. 

\noindent
From Lemma 3 and (\ref{prop_4})--(\ref{prop_5}), which are valid for all $\T(t)$, the profiles of $\FF_p$ in the elliptic/hyperbolic configuration are such that in every $\T(t)$ we have $\FF_p'=0$ when $\FF=\FF_p$, and so $\FF_p< \FF$ and $\FF_p'>0$ as $x$ increases asymptotically (see figure \ref{fig7}). We examine the sign of $\varphi$ in (\ref{phi_def}) for $\FF_p$ whose profile is as shown in figure \ref{fig7}. We consider first the case without a TV of $\Theta$. Given the properties of the curves of $\FF_p$ in all $\T(t)$, there will necessarily exist values $y<r_{\rm{b}}$ and $z>r_{\rm{b}}$ of the radial coordinate (see figure \ref{fig7}), such that $\FF(y)=\FF_p(y)$ and $\FF_p(z)=1$ hold. It is evident that the relation between $\FF_p(x)$ and $\FF_p(r)$ in (\ref{phi_def}) for a mixed elliptic/hyperbolic configuration strongly depends on the choice of the domain boundary. We use figure \ref{fig8} to illustrate how, given the curves of figure \ref{fig7}, the sign of $\varphi$ depends on the selected $\eta[r]$.

Following figure \ref{fig8}, it is evident that the elliptic/hyperbolic configuration allows for three types of possible domains $\eta[r]$ and behaviors of $\varphi$ in (\ref{phi_def}) along any $\T(t)$:
\bse\ba \fl {\rm{I}},\quad 0\leq r<y:\nonumber\\
\fl\qquad\FF_p(r)\leq \FF_p(x)\,\quad\forall\,x\in\eta[r],\qquad\quad\,\,\varphi\leq 0,\quad \C\leq 0\\
\fl {\rm{II}},\quad y<r<z:\quad \exists \,\,y_1\in\eta[r]\quad\hbox{such that}:\nonumber\\
\fl \qquad \FF_p(r)\leq \FF_p(x)\quad 0\leq x\leq y_1,\qquad\quad \varphi\leq 0,\\
\fl \qquad \FF_p(r)\geq \FF_p(x)\quad y_1\leq x\leq r,\qquad\quad \varphi\geq 0,\\
\fl {\rm{III}},\quad r>z:\nonumber\\
\fl \qquad \FF_p(r)\geq \FF_p(x) \quad \forall\,x\in\eta[r]\qquad\quad\,\, \varphi\geq 0,\quad \C\geq 0,\ea\ese
Hence, it is evident that $\C\geq 0$ holds for any domain $\eta[r]$ with $r\geq z$. Of course, $z$ will be different for different $\T(t)$, but given the fact that for all $\T(t)$ we have $\FF_p'>0$ for $x>y$, a value $z>r_{\rm{b}}$ fulfilling the desired result always exists in every $\T(t)$. 

The presence of a TV of $\Theta$ does not make a significant effect. If this TV occurs in domains along the inner or intermediary regions I and II:\, $0<x<z$, then its effect is the same as that discussed in proposition 7 and depicted in the bottom panel of figure \ref{fig5}. In this case, $C\geq 0$ holds (at least) in the more external part of $\eta[r]$ and has no effect on the hyperbolic region III, since $\Theta$ would already be monotonous in III, and so Proposition 2 would apply. If the TV of $\Theta$ occurs in the external zone III\, $x>z$, then we have exactly the same situation as in Proposition 3, and thus, irrespectively of the profile of $\Theta$, the result of this Proposition would apply: there always exist domains $\eta[r]$ with $r>z_1>z$ for which $\C\geq 0$ holds. As commented in the proof of Proposition 3, the presence of several TV's of $\Theta$ would not alter the result as long as there is a clear asymptotic behavior of $\Theta'$.    
\\

\noindent
{\underline{Note}}. 

\noindent
The main implication of Proposition 9 is that the determinant factor for $\C\geq 0$ is the existence of a regular hyperbolic asymptotic exterior, with local features (TV's of $\Theta$ or the enclosed elliptic region) playing a very minor role. In other words, given the non--local nature of back--reaction, local features can always be ``coarse grained'' when averaging domains are sufficiently large. As shown in figure \ref{fig8}, all domains of the form I are identical to elliptic domains without TV's, while all domains of the form III can be treated (as far as conditions for back--reaction are concerned) just like hyperbolic domains with or without TV's (of $\Theta$). Intermediary domains of the form II are transitional, being practically identical to elliptic domains with a TV of $\FF$ and $\Theta$.  

Notice that Proposition 9 is valid even if the internal elliptic region undergoes critical collapse conditions. If dust layers reach the collapsing singularity, then the $\T(t)$ after some $t$ will no longer be fully regular, becoming singular at the coordinates marking the singularity. The range $\eta[r]$ would be necessarily restricted. However, as explained in Appendix A, the function $\FF_p$ remains regular and this development simply requires one to treat the involved integrals as improper integrals, so that $\Theta_p,\,\Theta_q$ and other scalars are regular at all points save at the singularity locus.\\

\noindent
{\underline{Note: Spatial curvature in mixed configurations}}. 

\begin{figure}[htbp]
\begin{center}
\includegraphics[width=3in]{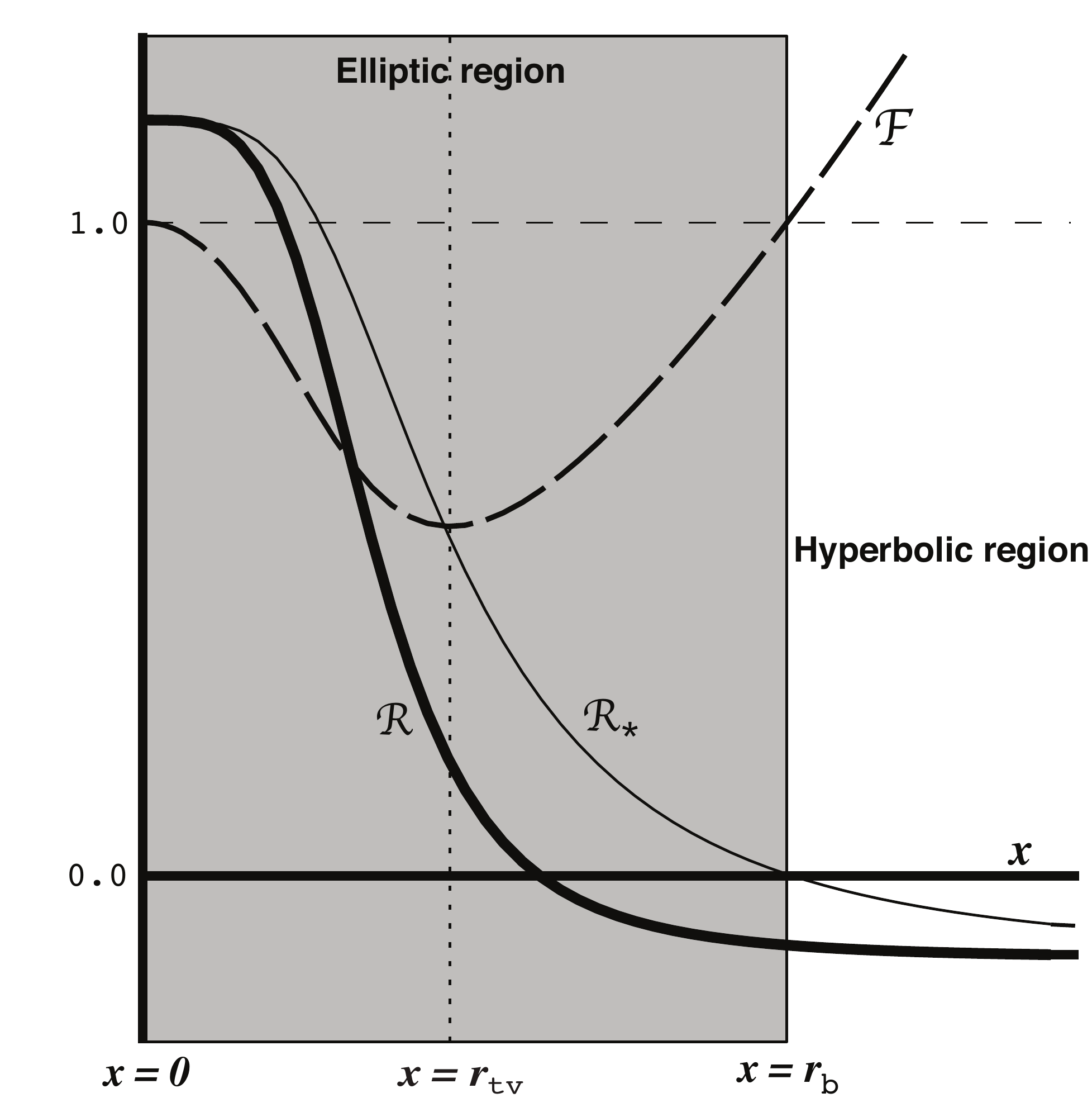}
\caption{{\bf Spatial curvature in an elliptic region surrounded by a hyperbolic exterior.} The figure displays the radial profile of the spatial curvature $\RR$ and dual quasi--local $\RR_q$ given by (\ref{qRR}). The function $\FF$ is diplayed for comparison. In order to have a positive $\RR$ near the center becoming negative at finite comoving coordinate values, there must be a sufficiently large negative gradient $\RR'$. In these conditions, $\RR\leq \RR_q$ and the critical value (\ref{critval}) for a TV of $\FF$: $\RR=\RR_q/3$ can be reached. At this value $\FF'=0$ and, as $x$ increases, $\FF$ increases until it reaches $\FF=1$ at the boundary $r=r_{\rm{b}}$ between the elliptic and hyperbolic regions. Notice that $\RR$ is already negative in areas inside the elliptic region where $\FF<1$ holds.   }
\label{fig9}
\end{center}
\end{figure}

\noindent
It is interesting to examine the behavior of the spatial curvature $\RR$ in ``mixed'' elliptic/hyperbolic regions of this type. Evidently, for positive curvature to become negative at a fixed comoving boundary, we must necessarily have a sufficiently large $\RR'<0$, which means that the critical value (\ref{critval}) is necessarily reached, as in elliptic domains with a TV of $\FF$ (Propositions 6 and 7). Since (from (\ref{qRR}) and (\ref{KTVa})) a zero of $1-\FF$ is a zero of the quasi--local curvature $\RR_q$ and for $\RR'\leq 0$ we have $\RR\leq \RR_q$, then (from (\ref{propq3})) condition $\FF(r_{\rm{b}})=1$ implies 
\begin{equation}  \RR(r_{\rm{b}})=\frac{1}{\VV_q(r_{\rm{b}})}\int_0^{r_{\rm{b}}}{\RR'\,\VV_q\,\dd x}<0.\end{equation}
Hence, for values near $r_{\rm{b}}$ (but $<r_{\rm{b}}$) in the elliptic region (for which $1-\FF^2$ is still positive) the local curvature is already negative. This is an example showing that an elliptic region ($1-\FF^2>0$) does not necessarily imply that $\RR>0$ holds in every point, though $\RR>0$ in every point does imply $1-\FF^2>0$. This situation is illustrated by figure \ref{fig9}.

\subsection{Parabolic exteriors.}

If instead of a hyperbolic exterior the elliptic region is surrounded by a parabolic exterior, we have 
\begin{equation}
\FF(x) \quad \left\{ \begin{array}{l}
   < 1,\quad 0 \le x < r_{\rm{b}}  \\ 
  = 1,\quad x \geq r_{\rm{b}}  \\  
 \end{array} \right.\label{Fprof2}
\end{equation}
It is evident that the replacement of the hyperbolic region by a parabolic one, simply removes the external zone (hyperbolic domains of type III) and keeps the intermediary zone (elliptic domains with a TV of $\FF$) all the way into the asymptotic range. 

\begin{quote}

{\underline{Proposition 10}}:\,\, In all regular LTB configurations made by a parabolic region surrounding an elliptic region containing a center, without a TV of $R$ but with possible TV's of $\Theta$, there are always domains $\eta[r]$ with $r>a$ for some $a>0$ for which $\C\geq 0$ holds. 
\end{quote}

\noindent
{\underline{Proof}}. 

\noindent
Since the behavior of $\FF$ and $\FF_p$ is qualitatively analogous to that of elliptic domains with TV's of $\FF$ (and possibly $\Theta$), the proof is afforded by Propositions 6 and 7.\\

\noindent
{\underline{Note}} 

\noindent
It is evident that the elliptic/parabolic configuration is qualitatively analogous to an elliptic model with the $\T(t)$ having topology $\mathbb{R}^3$ and spatial curvature decreasing ($\RR'<0$) and $\RR\to 0$ asymptotically. However, the assumption of the parabolic exterior ($\FF=1$ for $x\geq r_{\rm{b}}$) makes it much more artificial and contrived. The most interesting result is the fact that back--reaction can be positive in regular regions containing parabolic domains {\it {only}} if the latter do not contain a symmetry center.   

\begin{table}
\begin{center}
\begin{tabular}{|c| c| c| c|}
\hline
\hline
\hline
\hline
\multicolumn{4}{|c|}{Parabolic domains: $\FF=1$}
\\  
\hline
\hline
{Turning values} &{$Q$ and domain restrictions} &{Comments} &{$\ACal\geq 0$}
\\ 
\hline
\hline
{} &{} &{} &{}
\\
{None } &{$\QQ=0$ holds for all $\eta[r]$} &{Proposition 1} &{No}
\\
{(possibly $\Theta$)} &{} &{} &{}
\\
\hline
\hline  
\hline
\hline
\multicolumn{4}{|c|}{Hyperbolic domains: $\FF\geq 1$}
\\  
\hline
\hline
{Turning values} &{$Q$ and domain restrictions} &{Comments} &{$\ACal\geq0$} 
\\
\hline
\hline
{} &{} &{} &{}
\\
{None} &{$\QQ\geq 0$ holds for all $\eta[r]$} &{Proposition 2} &{Yes}
\\
{} &{} &{} &{} 
\\
{$\Theta$} &{$\QQ\geq 0$ holds for $\eta[r]$ with} &{Proposition 3} &{Yes}
\\
{} &{$0<r\leq r_1$ {\bf{or}} $r\geq r_2$ with $r_1<r_2$} &{} &{}
\\
{} &{} &{} &{}
\\   
\hline 
\hline
\hline
\hline
\multicolumn{4}{|c|}{Elliptic domains: 
\,$0\leq \FF\leq 1$}
\\  
\hline
\hline
{Turning values} &{$Q$ and domain restrictions} &{Comments} &{$\ACal\geq 0$}
\\ 
\hline
\hline
{} &{} &{} &{}
\\
{None} &{$\QQ\leq 0$ holds for all $\eta[r]$} &{Proposition 4} &{No}
\\
{} &{} &{} &{}
\\
{$\Theta$} &{$\QQ\geq 0$ holds for $\eta[r]$} &{Proposition 5} &{Extremely}
\\
{} &{with $0<r_1\leq r\leq r_2$} &{} &{restricted}
\\ 
{} &{} &{} &{}
\\ 
{$\FF$} &{$\QQ\geq 0$ holds for $\eta[r]$} &{Proposition 6} &{Yes}
\\ 
{} &{with $r\geq a$ with $a>0$} &{} &{}
\\ 
{} &{} &{} &{}
\\
{$\FF$ and $\Theta$} &{$\QQ\geq 0$ holds for $\eta[r]$ with} &{Proposition 7} &{Yes}
\\
{} &{$0<r\leq r_1$ {\bf{or}} $r\geq r_2$ with $r_1<r_2$} &{} &{}
\\
{} &{} &{} &{}
\\
{R} &{$\eta[r]$ with $r<\rtv$: as the cases above} &{Proposition 8} &{Extremely}
\\
{} &{$\eta[r]$ with $r>\rtv$: as with TV of $\FF$} &{Spherical topology} &{restricted}
\\
{} &{} &{} &{}
\\
\hline 
\hline
\hline
\hline
\multicolumn{4}{|c|}{Elliptic regions with a hyperbolic exterior}
\\  
\hline
\hline
{Turning values} &{$Q$ and domain restrictions} &{Comments} &{$\ACal \geq 0$}
\\ 
\hline
\hline
{} &{} &{} &{}
\\
{$\FF$ } &{$\QQ\geq 0$ holds for all $\eta[r]$} &{Proposition 9} &{Yes}
\\
{(possibly $\Theta$)} &{with $r\geq a$ for $a>0$} &{} &{}
\\
{} &{} &{} &{}
\\
\hline
\hline
\hline
\hline
\multicolumn{4}{|c|}{Elliptic regions with a parabolic exterior}
\\  
\hline
\hline
{Turning values} &{$Q$ and domain restrictions} &{Comments} &{$\ACal \geq 0$}
\\ 
\hline
\hline
{} &{} &{} &{}
\\
{$\FF$ } &{as above} &{Proposition 10} &{Yes}
\\
{(possibly $\Theta$)} &{} &{} &{}
\\
\hline
\hline
\end{tabular}
\end{center}
\caption{{\bf{Summary of sufficient conditions for back--reaction in LTB models}}. The domain $\eta[r]$ is defined by (\ref{etadef}). Turning values are defined in section 8. The parameters $r_1,\,r_2,\,\rtv$ and $a$ depend on the profiles of $\FF$ and $\Theta$ for each case and are given in the proofs of the Propositions. In elliptic domains the regularity conditions strongly constrain the profiles of $\FF$, and so conditions stated in the table might fail to hold (see sections 8--12 and Appendices A and B). Sufficient conditions for $\ACal\geq 0$ are discussed in the follow up paper (part II).}
\end{table}

\section{Final discussion and conclusion.}

We have provided in this article a fully comprehensive discussion of the back--reaction term $\QQ$ for LTB models in the context of Buchert's averaging formalism. All possible regular configurations admitting at least a symmetry center were examined in ten propositions that where rigorously proven for parabolic, hyperbolic and elliptic domains, regions and models, and for turning values (TV') of $\Theta,\,\FF$ and $R$ along hypersurfaces $\T(t)$. A summary of the conditions for $\QQ\geq 0$ is provided in Table 1. 

The form of $\C$ for all domains $\eta[r]$ leads to a determined sign of $\QQ$ for parabolic ($\QQ=0$ Proposition 1) and for hyperbolic ($\QQ\geq 0$) or elliptic ($\QQ\leq 0$) models without TV's (Propositions 2 and 4). In all other configurations, the sufficient conditions for the occurrence of $\QQ\geq 0$ depend on the selected domain $\eta[r]$ and on the existence of TV's of $\Theta$ or $\FF$. These conditions are not too restrictive for domains and models (hyperbolic or elliptic) whose hypersurfaces $\T(t)$ have topology $\mathbb{R}^3$ and have a full radial asymptotic range (Propositions 3, 5--7 and 9--10). For elliptic models whose $\T(t)$ have spherical topology (TV of $R$, Proposition 8), these conditions are much more restrictive.  

 From the proofs of the Propositions it is evident that the profiles of $\FF,\,\FF_p$ and their radial gradients determine the conditions for $\C\geq 0$.  However, as discussed in several parts of the article (for example, see equations (\ref{KTVa})--(\ref{KTVb}) in section 8.2, as well as sections 10 and 13), these radial profiles are themselves closely related to the spatial curvature $\RR$, its average quasi--local function $\RR_q$ and their gradients $\RR',\,\RR_q'$. Hence, statements about configurations favoring $\QQ\geq 0$ can also be given as statements on $\RR$ in specific domains. Thus, fulfillment of $\QQ\geq 0$ (and possibly $A_{\rm{eff}}\geq 0$) is compatible with the following domains
\begin{itemize}
\item $\RR\leq 0$ for all $r$ (``pure'' hyperbolic models of section 10).
\item $\RR\leq 0$ for large $r$, even if these domains contain elliptic inner regions with $\RR>0$ (hyperbolic models enclosing an elliptic region, section 13).
\item $\RR\geq 0$, with sufficiently large gradients $\RR'<0$ and $\RR\to 0$ asymptotically (elliptic domains with open topology and a TV of $\FF$, section 11)
\end{itemize}
This connection between the sign of $\QQ$ and the sign and profile of the spatial curvature (specially negative curvature) fits very well with the dynamical importance of this curvature due to the presence of large cosmic voids in the pattern of cosmic large scale structure~\cite{buch08}.

However, $\FF$ is an invariant scalar in LTB models~\cite{sussQL} and can also be associated with a covariant characterization (valid for spherical symmetry) of ``binding energy'', which has been proposed in \cite{sussQL,MSQLM,hayward1,hayward2,szab} by means of the comparison between the quasi--local or ``effective'' mass--energy function $\MM_q$ (the function $M$ in (\ref{fieldeq1})--(\ref{fieldeq2})) and the proper mass--energy $\MM_p$  function:
\bse\ba
2\MM_p(r) =\frac{2G}{c^2}\int_{\DD[r]}{\rho\,\dd\VV_p}=\kappa\int_0^r{\rho\,\FF^{-1}\,R^2\,R'\,\dd x},\\
2\MM_q(r) = 2M(r) =\frac{2G}{c^2}\int_{\DD[r]}{\rho\,\FF\dd\VV_p}=\kappa\int_0^r{\rho\,R^2\,R'\,\dd x},\ea\ese
where $\kappa=8\pi G/c^2$. A binding energy integral functional can be defined as
\ba\fl\B[r] &=& 2\left[\MM_q-\MM_p\right]=\frac{2G}{c^2}\int_{\DD[r]}{\rho\,(\FF-1)\,\dd\VV_p}=\kappa\int_0^{r}{\rho\,\frac{(\FF-1)\,R^2\,R'}{\FF}\,\dd x}\nonumber\\
\fl &=& \kappa\left\langle \left(\FF-1\right)\,\rho\right\rangle\, \VV,\label{Bdef}\ea
which for every domain is a conserved quantity along the 4--velocity flow: $\dot\B[r]=0$. Bearing in mind that the radial profiles of $\FF$ and $\FF_p$ are qualitatively analogous and that $\B[r]$ in (\ref{Bdef}) is not a local but an integral quantity associated to a whole domain $\eta[r]$, all statements about a given asymptotic behavior of $\FF_p$ can be given as statements on the behavior and profiles of this binding energy in different domains. Therefore, fulfillment of condition $\QQ\geq 0$ can be associated with domains in which
\begin{itemize}
\item $\B[r]\geq 0$ for all $r$ (``pure'' hyperbolic models of section 10).
\item $\B[r]\geq $ for large $r$, even if these domains contain elliptic inner regions with $\B[r]<0$ (hyperbolic models enclosing an elliptic region, section 13).
\item $\B[r]\leq 0$, where the gradient $\B'<0$ becomes less negative for large $r$ and $\B\to 0$ asymptotically (elliptic domains with open topology and a TV of $\FF$, section 11)
\end{itemize}
This connection between binding energy and back--reaction has been highlighted by Wiltshire~\cite{wiltshire2}.

It is important to notice that, regardless of the TV's of $\Theta$ or $\FF$, or changes of signs of spatial curvature or binding energy and other local complexities of the radial profiles, the conditions for positive $\C$ (sufficient for $\QQ\geq 0$) are basically determined by the asymptotic radial behavior of the incumbent scalars. The dominance of the asymptotic behavior is particularly evident in the case of the elliptic region with a hyperbolic exterior discussed in section 13 and illustrated by figures \ref{fig7} and \ref{fig8}: a local feature (like the elliptic region or a TV of $\Theta$) has no effect on the occurrence of $\C\geq 0$, as long as we select domains that lie sufficiently far away from the center. This is so, not only in the external hyperbolic region, but even in the transition zone still inside the elliptic region (see figure \ref{fig8}). The fact that the asymptotic behavior is the key factor is expected, as back--reaction is a non--local effect, and so local features can always be ``coarse grained'', even if they exhibit critical conditions like a collapsing singularity in the elliptic region.     
     
The Propositions that we have proved provide information on the existence of a positive $\QQ$, but not on how large this term can be.  It is possible to argue qualitatively that, while  back--reaction can be positive in elliptic and hyperbolic domains, the latter are likely to yield a larger value. This follows from comparing the integrands of $\Phi$ and $\Psi$ in (\ref{CC2})--(\ref{psi_def}) for each type of domain. Assuming the gradient $\Theta'$ to be of similar magnitude in each case, the function $\varphi$ in (\ref{phi_def}) is likely to be a small quantity in most of the asymptotic radial range of an elliptic domain compatible with $\C>0$. If we look at the profile of $\FF_p$ displayed by figure \ref{fig4}: as $r/y\gg 1$, we have $\FF\to 1$ but $\FF<1$ for all $x$. Since $\FF_p<\FF$, then $\FF_p\to 1$ (asymptotically small curvature). But then, $\FF_p(x)\approx\FF_p(r)\approx 1$, which makes $\varphi$ in (\ref{phi_def}) close to zero for a long asymptotic radial range (notice also that $\FF$ and $\FF'$ are strongly constrained by (\ref{FFell1})). In comparison, $\FF_p$ is not restricted to remain below $\FF_p=1$ in the asymptotic range of hyperbolic domains or regions (see figures \ref{fig7} and \ref{fig8}), and so this function can reach much larger values than in elliptic regions, which makes the integrand of $\Phi$ much less restricted.   

The magnitude of the back--reaction term $\QQ$ is also related to the magnitude of the radial gradients of various scalars. Using (\ref{Vprime}), (\ref{prop2}), (\ref{Vqprime}), (\ref{propq2}), (\ref{CCdef}) we can write $\C$ as
\begin{equation}\fl \C = \left(\frac{R}{3R'}\right)^2\,\left[\left(\Theta_p'\frac{\FF}{\FF_p}\right)^2-\Theta_q'{}^2\right]=\left(\frac{R}{3R'}\right)^2\,\left[\Theta_p'\frac{\FF}{\FF_p}-\Theta_q'\right]\left[\Theta_p'\frac{\FF}{\FF_p}+\Theta_q'\right].\label{CCgrads}\end{equation}
However, the gradients $\Theta_p'$ and $\Theta_q'$ can be related by means of (\ref{Hamcon}) and (\ref{ave_eveqs_ef3}) to the gradients of $\rho_p,\,\rhoav,\,\RR_q$ and $\RRav$, while gradients of $\RR_q$ relate to gradients of $\FF$, then it is evident that $\C$ will be larger in radial ranges where these gradients are significant ({\it{i.e.}} in regions where inhomogeneity is not negligible). As a consequence of (\ref{CCgrads}) and the qualitative arguments in the previous paragraph, it seems that hyperbolic models that allow for larger gradients of $\FF$ and $\FF_p$ (in a wider radial range) are much more likely to provide a back--reaction $\QQ$ that is sufficiently large for $\ACal>0$ in (\ref{efeacc}) to hold for density profiles that are not too restrictive, though  even if the resulting $\C>0$ could be small in elliptic models an effective acceleration could still occur if $\rho$ is sufficiently small.     

Since $\Theta_p$ (or $\Thetaav$) and $\Theta_q'$ in (\ref{Hamcon}) and (\ref{ave_eveqs_ef3}) play the role of ``kinetic energy'' terms, the relation between $\C$  and the gradients $\Theta_p'$ and $\Theta_q'$ given by (\ref{CCgrads}) provides theoretical and contextual support to Wiltshire's interpretation of back--reaction~\cite{wiltshire2}. Given the LTB configurations which we have proven to be compatible with $\QQ\geq 0$, we examine in a follow up paper (part II) the magnitude and radial variation of the back--reaction term, as well as boundary conditions that are needed to find the conditions for the fulfillment of (\ref{efeacc}) and/or (\ref{efeacc2}), which lead to the existence of a positive effective acceleration that could mimic the effects of dark energy.  Another line for future work is to extend the results presented here and in part II to the more general Szekeres models~\cite{kras,bolejko1}. These tasks are being undertaken in separate continuing articles.

\appendix
\section{Regularity of LTB models.}

Standard regularity conditions of LTB models are essential for validity of the results presented in this artice. These conditions have been extensively discussed in the literature~\cite{kras,ltbstuff1,ltbstuff2,ltbstuff3,ltbstuff,suss02}. We provide here a brief summary. 

\subsection*{Symmetry centers}

We have only considered LTB models having (at least) one symmetry center, which is a regular timelike  worldline corresponding to a fixed point of the rotation group SO(3). This is a sufficient condition for integrals in (\ref{ave_def}) and (\ref{aveq_def}) to be finite in a domain $\eta[r]$ defined by (\ref{etadef}). The symmetry center can be marked as $r=0$ (and $r=r_c$ if there is a second one). The following conditions hold: $R(t,0)=\dot R(t,0)=0$, while $\FF(0)=1$ and $M(0)=M'(0)=\FF'(0)=0$, but $R'\to 1$ as $r\to 0$. Also, for any scalar $A$ and its dual functions $A_p$ and $A_q$ obtained by means of (\ref{ave_def}) and (\ref{aveq_def}) we have  
\bse\ba A(t,0)=\Aav[0]=A_p(t,0)=A_q(t,0),\\ A'(t,0)=\Aav'[0]=A_p'(t,0)=A_q'(t,0)=0,\ea\ese
From (\ref{Cdef}), (\ref{CC1}) and the equations above, it is evident that back--reaction vanishes at the center: $\C(0)=\C'(0)=0$. Also, we have from (\ref{Edef}):\, $\ACal_*(0,r)=-(\kappa/2)\rho(0)<0$, hence there is a nonzero  effective deceleration in all domains sufficiently closed to a center in all LTB models for which $\rho(0)>0$. Notice that a central singularity is also associated with $R=0$, but its coordinate locus is not a comoving worldline.

\subsection*{Regularity at a TV of $R$ and shell crossing singularities.}

There is a ``shell--crossing'' singularity if $R'(t,r)=0$ occurs for coordinates that are not of the type $r=$constant, thus an important regularity condition is then~\cite{ltbstuff1,ltbstuff3,ltbstuff,suss02}
\begin{equation} R'(t,r) > 0,\label{noshx}\end{equation}
holds with $R'=0$ occuring (regularly) only in models whose $\T(t)$ have spherical topology at a TV of $R$ (see Appendix B). 

Necessary and sufficient conditions to avoid shell crossing singularities can be given in full analytic form. This is done in \cite{ltbstuff1,ltbstuff3,ltbstuff} in terms of the free functions associated with (\ref{fieldeq1}) and using the function $E$ related to $\FF$ by $\FF=[1+E]^{1/2}$, while in \cite{suss02} it is done on initial conditions specified on a fiducial $\T(t_i)$, and using $K$ defined by $\FF=[1-K]^{1/2}$. 

In hyperbolic domains and models the conditions to avoid shell crossings necessarily require $M'$ and $\FF'$ to have the same sign as $R'$, which prevents the existence of TV's of $\FF$ (see equation (22) of \cite{ltbstuff}). However, these regularity conditions allow for a  TV of $\FF$ in elliptic models, though the gradients $\FF'$ are strongly constrained, as shown by equation (23) of \cite{ltbstuff} and equation (76) of \cite{suss02}. It is useful to rewrite the latter equation in terms of $\FF$ and its gradients, $M$ and its gradients and $R_i=R(t_i,r)$ where $t_i$ marks an arbitrary initial $\T$:
\begin{equation}\fl 2\pi P_i\left[\delta_i^{(m)}-\frac{3}{2}\delta_i^{(k)}\right]\geq P_iQ_i\left[\delta_i^{(m)}-\frac{3}{2}\delta_i^{(k)}\right]-\left[\delta_i^{(m)}-\delta_i^{(k)}\right]\geq 0,\label{FFell1}\end{equation} 
where
\bse\ba \fl \delta_i^{(m)} \equiv \frac{M'/M}{3R_i'/R_i}-1,\qquad \frac{3}{2}\delta_i^{(k)}=-\frac{\FF\FF'}{(1-\FF^2)R_i'/R_i}-1,\label{FFell2a}\\
\fl P_i\equiv \pm \frac{[1-(1-v)^2]^{1/2}}{v^2},\qquad Q_i\equiv {\rm{arccos}}(1-v)-[1-(1-v)^2]^{1/2},\label{FFell2b}\\
\fl v \equiv \frac{(1-\FF^2)\,R_i}{M}=\frac{[\RR_q]_i}{\kappa[\rho_q]_i}.\label{FFell2c}\ea\ese
Thus, avoidance of shell crossings places much stronger constraints on the radial profile of $\FF$ in elliptic models than in hyperbolic or parabolic models. This fact is important for the discussion of sections 10 and 11.

\subsection*{Range restrictions due to a collapsing singularity.}

So far we have assumed that the integration range $\eta[r]$ defined by (\ref{etadef}) is fully regular. However, it is a well known fact that a collapsing singularity arises in elliptic LTB models when $\Theta<0$, and the coordinate locus of this singularity is (in general) not simultaneous ({\it i.e.} not marked by a constant $t$ or single $\T(t)$). In general, this collapsing singularity is marked by a curve $[t(r_{\rm{coll}}), r_{\rm{coll}}]$ in the $(t,x)$ coordinate plane, where $R(t(r_{\rm{coll}}), r_{\rm{coll}})=0$ and curvature scalars diverge (see \cite{suss08}). Hence, in any collapsing elliptic region the hypersurfaces $\T(t)$ for $t\geq t(r_{\rm{coll}})$ are only regular for the semi open subset $\bar\eta[r]\equiv \{x\,|\, r_{\rm{coll}} < x \leq r\} \subset \eta[r]$. However, the existence of this singularity has no consequence in the definition of $\Aav[r]$ or $A_p$ and $A_q$ functions because the involved integrals can be treated simply as standard improper integrals. We define at each $\T(t)$ the incumbent integrals with their lower integration limit as $y =r_{\rm{coll}}+\epsilon$, for an arbitrarily small $\epsilon>0$, and then obtain the limit as $\epsilon\to 0$. Off course, since $\Theta\to -\infty$ in this limit, $\Thetaav$ or $\Theta_p$ or $\Theta_q$ might diverge as well, but the functions are well defined and behaved in the range $\bar\eta[r]$.  Regarding $\FF_p=\VV_q/\VV_p$, the integrals involved it its definition are regular as $x\to r_{\rm{coll}}$. Therefore, the restrictions mentioned above simply prevent $\FF_p$ from taking values $x<r_{\rm{coll}}$, and so the results presented in sections 7-9 for any configuration that involves collapsing layers can be trivially extended to include hypersurfaces $\T(t)$ for $t\geq t(r_{\rm{coll}})$.

\section{Radial dependence and proper length.}

While spherical symmetry effectively reduces the spatial variation of scalars on the $\T(t)$ to a one--dimensional radial dependence, there is no inherent covariant meaning in the radial coordinate (in fact, the metrics (\ref{LTB1}) and (\ref{spmetric}) are invariant under arbitrary re--scalings $r=r(\bar r)$). However, radial rays are totally geodesic (they are spacelike geodesics of (\ref{LTB1})), so it is possible and desirable to relate radial dependence at each $\T(t)$ to dependence on the natural (and covariant) parametrization of the rays in terms of their affine parameter or proper length:
\begin{equation} \xi(r)=\int_{0}^{r}{\sqrt{g_{rr}}\dd x}=\int_0^r{\frac{R'}{\FF}\dd x},\label{xidef}\end{equation}
so that $\xi(0)=0$, and we have adopted the notation $\int_0^r{}=\int_{x=0}^{x=r}{}$, so that (unless specified otherwise) all functions inside an integral depend on the dummy variable $x$ (and as stated before, $t$ is a fixed parameter).

Since $\xi$ is a proper length, we must necessarily have $\xi(r)>0$ and $\xi'(r)>0$ for $r>0$, so that $r_2>r_1$ implies $\xi(r_2)>\xi(r_1)$ and the converse is also true. These properties define through (\ref{xidef}) a well behaved radial coordinate as that complying at every $\T(t)$ with the following regularity condition  
\begin{equation}\hbox{sign}\,(R') = \hbox{sign}\,\FF,
\qquad\,\, \exists\;\;\hbox{TV of}\, R \quad\Leftrightarrow\quad \exists\;\;\hbox{zero of}\,\,\FF,\label{RrF}\end{equation}
where ``TV of $R$'' stands for a regular ``turning value'' of $R$, or a zero of the gradient $R'$ under regular conditions (see section 8).~\footnote{This definition obviously excludes shell crossing singularities discussed in the previous Appendix for which $R'=0$ occurs in violation of (\ref{RrF}).}    

The relation between profiles of scalars with respect to $r$ and of $\xi$ at each $\T$ is given by
\bse\ba A' = \frac{\partial A}{\partial \xi}\,\xi'=\frac{\partial A}{\partial \xi}\,\frac{R'}{\FF}\label{Arxi}\\
 R' = \frac{\partial R}{\partial \xi}\,\xi'=\frac{\partial R}{\partial \xi}\,\frac{R'}{\FF},\quad\Rightarrow\quad \FF(r(\xi)) = \frac{\partial R}{\partial \xi},\label{Rrxi}\ea\ese
which (as long as (\ref{RrF}) holds) implies that
\bse\ba  
\fl \qquad\hbox{sign}\,(A') = \hbox{sign}\,(\partial A/\partial\xi),
\qquad \exists\;\;\hbox{TV of}\, A(r) \quad\Leftrightarrow\quad \exists\;\;\hbox{TV of}\,\, A(\xi),\label{sz1}\\
\fl \qquad\hbox{sign}\,(R') = \hbox{sign}\,(\partial R/\partial\xi),
\qquad  \exists\;\;\hbox{TV of}\, R(r) \quad\Leftrightarrow\quad \exists\;\;\hbox{TV of}\,\, R(\xi),\label{sz2}\ea\ese
so that radial dependence is qualitatively analogous to proper length dependence at each $\T(t)$: monotonous radial profiles $A(r)$ correspond to a monotonous profiles $A(\xi)$ and a TV of $A$ corresponds to a zero of $\partial A/\partial\xi$. Notice that (\ref{Rrxi}) allows us to relate $\FF$ to initial value $R_i=R(t_i,r)$ at at a fiducial hypersurface $\T_i$. Since this relation between $\FF$ and $\partial R/\partial\xi$ is valid in all $\T(t)$, a zero of $\FF$ (characteristic of $\mathbb{S}^3$ topology) or a monotonous profile of $\FF$ (characteristic of $\mathbb{R}^3$ topology) will be common to all $\T(t)$.

\section{The fluid--flow evolution equations.}

LTB models are usually examined by means of the solutions of the Friedman--like equation (\ref{fieldeq1}). Since these models (as all spherically symmetric spacetimes) can be completely described by covariant scalars~\cite{LRS}, an alternative approach to study their dynamics is through the evolution equations for the scalars 
 \begin{equation} A=\{\rho,\,\Theta,\,\Sigma,\,\EE,\,\RR\},\label{cov_scals}\end{equation}
which follow from equations (\ref{fieldeq2}) and (\ref{Theta1})--(\ref{EE1}) and provide a complete set to characterize the LTB models. This is the ``fluid flow'' or covariant ``1+3'' framework of Ehlers, Ellis, Bruni, Dunsbury and van Ellst~\cite{ellisbruni89,BDE,1plus3,LRS} and leads to the evolution equations    
\bse\ba
\dot\Theta &=&-\frac{\Theta^2}{3}
-\frac{\kappa}{2}\,\rho-6\Sigma^2,\label{ev_theta_13}\\
\dot \rho &=& -\rho\,\Theta,\label{ev_mu_13}\\
\dot\Sigma &=& -\frac{2\Theta}{3}\,\Sigma+\Sigma^2-\EE,
\label{ev_Sigma_13}\\
 \dot\EE &=& -\frac{\kappa}{2}\rho\,\Sigma
-3\,\EE\,\left(\frac{\Theta}{3}+\Sigma\right),\label{ev_EE_13}\ea\ese
together with the spacelike constraints  
\begin{equation}
\left(\Sigma+\frac{\Theta}{3}\right)'+3\,\Sigma\,\frac{R'}{R}=0,\qquad
\frac{\kappa}{6}\rho\,'
+\EE\,'+3\,\EE\,\frac{R'}{R}=0,\label{ccons_13}\end{equation}
and the ``Hamiltonian'' constraint (analogous to the Friedman equation)
\begin{equation}\left(\frac{\Theta}{3}\right)^2 = \frac{\kappa}{3}\, \rho
-\frac{\RR}{6}+\Sigma^2,\label{cHam_13}\end{equation}
which is an integral of the Raychaudhuri equation (\ref{ev_theta_13}).

Buchert's equations (\ref{ave_eveqs_ef1})--(\ref{QQgen}) follow directly by applying the proper average functional (\ref{ave_def}) to both sides of equations (\ref{ev_theta_13}), (\ref{ev_mu_13}) and (\ref{cHam_13}) (the Raychaudhuri and energy balance equations and Hamiltonian constraint), and using the time derivative rule (\ref{tconm}), as well as the formulae for the covariance and variance momenta (\ref{var}) and (\ref{covar}). The integrability condition
\begin{equation} \dot\QQ +2\Thetaav\,\QQ+\frac{2}{3}\Thetaav\,\RRav+\RRav\,\dot{}=0,\label{consistQR}\end{equation}
can be proven to be compatible with the fluid flow evolution equations (\ref{ev_theta_13})--(\ref{ev_EE_13}) and the Hamiltonian constraint (\ref{cHam_13}). The proof follows after a very long algebraic manipulation expressing  $\Sigma$ and $\EE$ as in (\ref{Sigma2})--(\ref{EE2}), together with applying (\ref{tconm}), (\ref{var}) and (\ref{covar}).   

\section{Proofs of Propositions 7 and 8} 

\subsection{Proposition 7.}

Consider an elliptic domain with a TV of $\FF$ and a TV of $\Theta$ (but not of $R$). The profiles of $\Theta'\varphi$ and $\Theta'\psi$ for this configuration are those displayed in the bottom panel of figure \ref{fig5}), with $x=y$ marking the location of the sign change of $\varphi$ and $x=\rtv$ marking the TV of $\Theta$. This leads to
\bse\ba \Phi=-\int_0^y{\Theta'|\varphi|\,\dd{x}}+\int_y^{\rtv}{\Theta'\varphi\,\dd x}-\int_{\rtv}^r{|\Theta'|\varphi\,\dd x},\label{AC1a}\\
\Psi=\int_0^{\rtv}{\Theta'\psi\,\dd x}-\int_{\rtv}^r{|\Theta'|\psi\,\dd x}.,\label{AC1b}\ea\ese
We introduce now the real positive numbers $\{\alpha,\,\beta,\,\gamma,\,\delta,\,\epsilon\}$ by
\bse\ba  0\leq \int_0^y{\Theta'|\varphi|\,\dd{x}}\leq \alpha\,y,\qquad 0\leq \int_y^{\rtv}{\Theta'\varphi\,\dd x}\leq \beta(\rtv-y),\label{AC2a}\\  0\leq \int_{\rtv}^r{|\Theta'|\varphi\,\dd x}\leq \gamma(r-\rtv), \qquad 0\leq \int_0^{\rtv}{\Theta'\psi\,\dd x} \leq \delta\rtv,\label{AC2b}\\  0\leq \int_{\rtv}^r{|\Theta'|\psi\,\dd x}\leq \epsilon(r-\rtv).\label{AC2c}\ea\ese
Condition $\C=\Phi\Phi\geq 0$ implies
\begin{equation} \frac{r}{\rtv}\leq \hbox{min}\left\{1-\frac{\alpha+\beta}{\gamma}\frac{y}{\rtv}+\frac{\beta}{\gamma},\,1+\frac{\delta}{\epsilon}\right\}\label{AC3}\end{equation}
or
\begin{equation} \frac{r}{\rtv}\geq \hbox{max}\left\{1-\frac{\alpha+\beta}{\gamma}\frac{y}{\rtv}+\frac{\beta}{\gamma},\,1+\frac{\delta}{\epsilon}\right\}.\label{AC4}\end{equation}
which provides the values of the real numbers $r_1$ and $r_2$. 

\subsection{Proposition 8.}

The profiles of  $\Theta'\varphi$ and $\Theta'\psi$ associated with Proposition 8 are qualitatively analogous with those of figure \ref{fig5}, and so the proof above applies also to that case. The only difference is that the role of $y$ above corresponds to the value $z$ such that $\FF_p(z)=|\FF_p(r)|$, which marks a zero of $\psi$ where this function passes from negative to positive (see (\ref{topS31})--(\ref{topS32}) in section 12). However, since $r/\rtv$ cannot take arbitrary large values, the fulfillment of condition (\ref{AC4}) is much more restrictive.   
\\

\section{Proof that a TV of $R$ must be a common TV of $\Theta$.}

A TV of $R$ ({\it{i.e.}} $R'(\rtv)=0$ for $\rtv\in\eta[r]$) occurs in regular conditions only in elliptic models whose $\T(t)$ are homeomorphic to $\mathbb{S}^3$. We prove in this section that 
\begin{equation} R'(\rtv)=0\quad\Rightarrow\quad \Theta'(\rtv)=0, \label{TVRT}\end{equation}
and show (as part of the proof) that this implication also holds for all quasi--local scalars $A_q$ obtained from (\ref{aveq_def}), as well as for the scalars $A=\rho,\,\RR,\,M$, but not for $\FF,\,\FF_p$ and scalars $A_p$ that follow from (\ref{ave_def}). Notice that the converse statement of (\ref{TVRT}) is false: there can be a TV of $\Theta$ with $R'>0$. 

From the regularity condition (\ref{RrF}), a well defined proper radial length (\ref{xidef}) requires the ratio $R'/\FF$ to be continuous (at least $C^0$) at $r=\rtv$. Since $R'(\rtv)=\FF(\rtv)=0$, with $R''(\rtv)<0,\,\FF'(\rtv)<0$, then from l'H\^opital rule we have at the limit $r\to\rtv$
\begin{equation}  \frac{R'}{\FF}\to \frac{R''(\rtv)}{\FF'(\rtv)}>0, \label{RrFtv}\end{equation}
which implies that $\rtv$ is also a TV of the quasi--local volume $\VV_q$ (because $\VV_q'=4\pi R^2R'$), but not of the proper volume $\VV_p$ (because $\VV_p'=4\pi R^2R'/\FF$). By considering (\ref{RrFtv}), together with (\ref{Vprime}), (\ref{prop2}) and (\ref{propq2}):
\begin{equation} A_q'=\frac{3R'}{R}\left[A-A_q\right],\qquad A_p'=\frac{3R'\,\FF_p}{R\,\FF}\left[A-A_p\right],\label{Apq_grads}\end{equation}
it is evident that for all $A_q$
\begin{equation} R'(\rtv)=0\quad\Rightarrow\quad A_q'(\rtv)=0, \label{TVRTq}\end{equation}
but, because of (\ref{RrFtv}), $\rtv$ is (in general) not a TV of the p--scalars $A_p$.

To prove (\ref{TVRT}) we assume that $\rtv$ is a TV of $R$ and examine the behavior of $\Theta$ and $\Theta_q$ around $r=\rtv+\epsilon$ for $|\epsilon|\ll 1$. Considering (from (\ref{TVRTq})) that a TV of $R$ implies $\Theta_q'(\rtv)=0$, we have at leading orders: 
\begin{equation}\fl \Theta_q(\rtv+\epsilon)\approx \Theta_q(\rtv) +\frac{1}{2}\Theta_q''(\rtv) \epsilon^2,\quad R(\rtv+\epsilon)\approx R(\rtv) +\frac{1}{2}R''(\rtv) \epsilon^2, \label{TRappr}\end{equation}
Since, in general, $\Theta(\rtv)\ne 0$, by applying (\ref{Apq_grads}) to $\Theta$ and $\Theta_q$ and at leading orders we get
\begin{equation} \fl\Theta(\rtv+\epsilon)-\Theta(\rtv)=\Theta_q(\rtv+\epsilon)-\Theta_q(\rtv)+\left[\frac{\Theta_q' R}{3R'}\right]_{\rtv+\epsilon}-\left[\frac{\Theta_q' R}{3R'}\right]_{\rtv}.\end{equation}
Using (\ref{TRappr}) to evaluate this expression at leading orders and dividing both sides by $\epsilon$ we get
\begin{equation} \frac{\Theta(\rtv+\epsilon)-\Theta(\rtv)}{\epsilon}=\frac{\Theta_q(\rtv+\epsilon)-\Theta_q(\rtv)}{\epsilon}+\frac{1}{6}\,\Theta_q''(\rtv)\,\epsilon.\end{equation}
Taking in both sides the limit as $\epsilon\to 0$ we get $\Theta'(\rtv)=\Theta_q'(\rtv)=0$, which is the desired result. This proof can be extended to other scalars (except $\FF$ which necessarily vanishes at $\rtv$).

\end{document}